\documentclass{PoS}

\usepackage{graphicx}
\usepackage{multirow}
\usepackage{arydshln}
\usepackage{slashed}

\def\OMIT#1{{}}
\def\lsim{\mathrel{\!\mathpalette\vereq<}\!}
\def\gsim{\mathrel{\!\mathpalette\vereq>}\!}
\def\vereq#1#2{\lower3.5pt\vbox{\baselineskip1.5pt \lineskip1.5pt
\ialign{$#1\hfill##\hfil$\crcr#2\crcr\sim\crcr}}}

\newcommand{\beq}{\begin{equation}}
\newcommand{\eeq}{\end{equation}}
\newcommand{\beqa}{\begin{eqnarray}}
\newcommand{\eeqa}{\end{eqnarray}}
\newcommand{\TeV}{{\rm TeV}}
\newcommand{\GeV}{{\rm GeV}}

\def\rhobar{\bar\rho}
\def\etabar{\bar\eta}
\def\lqcd{\Lambda_{\rm QCD}}

\newcommand{\Bbar}{\,\overline{\!B}{}}
\newcommand{\Dbar}{\,\overline{\!D}{}}
\newcommand{\Kbar}{\,\overline{\!K}{}}
\def\B0bar{\Bbar{}^0}
\def\D0bar{\Dbar{}^0}
\def\K0bar{\Kbar{}^0}

\newcommand{\Jpsi}{J\!/\!\psi}
\newcommand{\Gam}[2]{\bar{\Gamma}(#1 \! \to \Jpsi\, #2)}

\title{Flavor Constraints on New Physics}
\ShortTitle{Flavor Constraints on New Physics}

\author{\speaker{Zoltan Ligeti}\\
Lawrence Berkeley National Laboratory,
University of California, Berkeley, CA 94720, USA\\
E-mail: \email{ligeti@berkeley.edu}}

\abstract{This talk highlights, from a theoretical point of view, some recent
exciting results in flavor physics, as well as future prospects.  We discuss
possible implications of a subset of the experimental results in tension with
the standard model, such as the $4\sigma$ deviation in the $B\to
D^{(*)}\tau\bar\nu$ decay rates, and recent improvements in the constraints on
axion portal dark matter models.  We use the examples of constraining new
physics contributions to neutral meson mixing and the search for possible
vector-like fermions to illustrate the expected progress over the next decade to
increase the sensitivity to new physics at shorter distance scales.  We also
speculate about the ultimate limitations of (quark) flavor physics probes of new
physics.}

\FullConference{XXVII International Symposium on Lepton Photon Interactions at
High Energies\\
17-22 August 2015\\
Ljubljana, Slovenia}

\begin{document}

\section{Introduction}

I was asked to talk about flavor physics constraints on new physics (NP).  A
slight complication is that in the absence of unambiguous observations of
deviations from the standard model (SM) in laboratory experiments so far, and
with the LHC pushing the scale of NP higher, there are not really good
``simplified models" for non-SM flavor physics, containing a modest number of
parameters, which a large class of NP models match onto.  There are many flavor
physics constraints, and there are many NP models, and attempts to simplify this
large (and for most people uninspiring) matrix of constraints has only achieved
limited success for a handful of processes.  In other words, the interesting
information from flavor physics is not simple (as it depends on a large number
of processes and the theory is often complicated), and the simple information
from flavor physics is not interesting; for example, we learn little from just
the values of Cabibbo-Kobayashi-Maskawa (CKM) matrix elements\footnote{As
Lincoln Wolfenstein sometimes said, even though he invented the Wolfenstein
parameters, he did not care what their values were, only whether many
overconstraining measurements gave consistent determinations.}.

Flavor physics is the study of interactions that distinguish the three
generations of fermions, i.e., interactions that break the global $[U(3)]^5$
symmetry of the SM, of which each $U(3)$ acts on one of the 5 fermion
representations ($Q$, $u$ $d$, $L$, $e$).  In the SM, this symmetry is broken by
the Yukawa couplings, while in the presence of new physics there generally are
additional sources of flavor (and $CP$) violation.  We do not understand the
flavor structure of the SM, and if there is new physics at the 1--100\,TeV
scale, we have to understand the mechanism that suppresses its effects in the
flavor sector to satisfy the experimental constraints.  In addition, the
observed baryon asymmetry of the Universe requires $CP$ violation beyond the SM
(although that need not occur in the quark sector, nor necessarily in flavor
changing processes).  In any case, flavor measurements provide rich and
sensitive ways to probe the SM and search for NP, and flavor physics strongly
constrains any NP within the LHC reach.  These measurements will reach much
higher sensitivities in the next decade. 

The sensitivity of flavor measurements to very high mass scales typically comes
from large SM suppressions.  As a simple (and historically important) example,
consider $K^0$\,--\,$\K0bar$ mixing.  The splitting between the two mass
eigenstates is $\Delta m_K/m_K \sim 7 \times 10^{-15}$.  In the SM, $\Delta m_K$
arises dominantly from box diagrams with virtual $W$ bosons and $c$ quarks, and
can be estimated as
\beq
\frac{\Delta m_K}{m_K} \sim \alpha_w^2\, |V_{cs}V_{cd}|^2\,
  \frac{m_c^2}{m_W^4}\, f_K^2\,.
\eeq
The result is suppressed by CKM angles, a loop factor, the weak coupling, and
the GIM mechanism.  If a heavy particle, $X$, with effective $\bar s d X$
coupling, $g$, contributes an ${\cal O}(1)$ fraction to $\Delta m_K$, then
\beq
\bigg|{\Delta m_K^{(X)}\over \Delta m_K^{\rm (exp)}}\bigg|
  \sim \bigg|{g^2\, \lqcd^3\over M_X^2\, \Delta m_K^{\rm (exp)}}\bigg|
  \quad\Rightarrow\quad \frac{M_X}g \gtrsim\, 2\times 10^3\, \TeV\,.
\eeq
So even TeV-scale particles with loop-suppressed couplings $[g \sim {\cal
O}(10^{-3})]$ can give observable effects.  Thus, flavor measurements probe the
TeV scale if the NP has SM-like flavor structure, and much higher scales if the
NP flavor structure is generic.  In the SM only $W$ bosons change fermion
flavor, so flavor-changing processes of the known fermions (except the $t$
quark) are suppressed~by~at~least the second power of a high scale ($G_F
\propto 1/m_W^2$ in the SM, or $1/M_{\rm NP}^2$) and often by additional small
coefficients.  We want to find out if the higher dimension operators generated
by the high-scale physics have coefficients as predicted by the SM, and if
operators forbidden in the SM (e.g., right handed currents) are generated.

While new physics has been widely expected to occur at the TeV scale, hinted by
the hierarchy problem and the WIMP paradigm, after the discovery of the Higgs
boson, no other new particle is guaranteed to be observable in near future
laboratory experiments.  In flavor physics, typically kaons probe the highest
scales, since the SM suppressions are strongest for flavor-changing neutral
currents (FCNC) between 1st and 2nd generation quarks.  In many NP scenarios the
3rd generation is rather different from the first two, so there is strong
motivation to explore what the technology allows us to probe.  I find it fairly
certain that if new physics is discovered, we will eventually understand it as
``natural", no matter how ``strange" it might seem at first.

Section~\ref{sec:status} summarizes the current status of (quark) flavor physics
and reviews some tensions with the SM predictions.  These are some of the most
often discussed topics recently, and they are also interesting because they may
have the best chance to be established as clear deviations from the SM, as more
data is accumulated.  (I include four recent
measurements~\cite{Abdesselam:2016cgx, Abdesselam:2016llu, Aaboud:2016ire,
Aaij:2016yze} for completeness, which appeared since the conference.) 
Section~\ref{sec:future} gives some examples of the expected future progress and
improvements in sensitivity to NP, independent of the current data. 
Section~\ref{sec:concl} contains some comments on the ultimate sensitivity of
flavor physics experiments to NP.

\section{Status of flavor physics}
\label{sec:status}

\begin{figure}[t]
\centerline{\includegraphics[width=.56\textwidth]{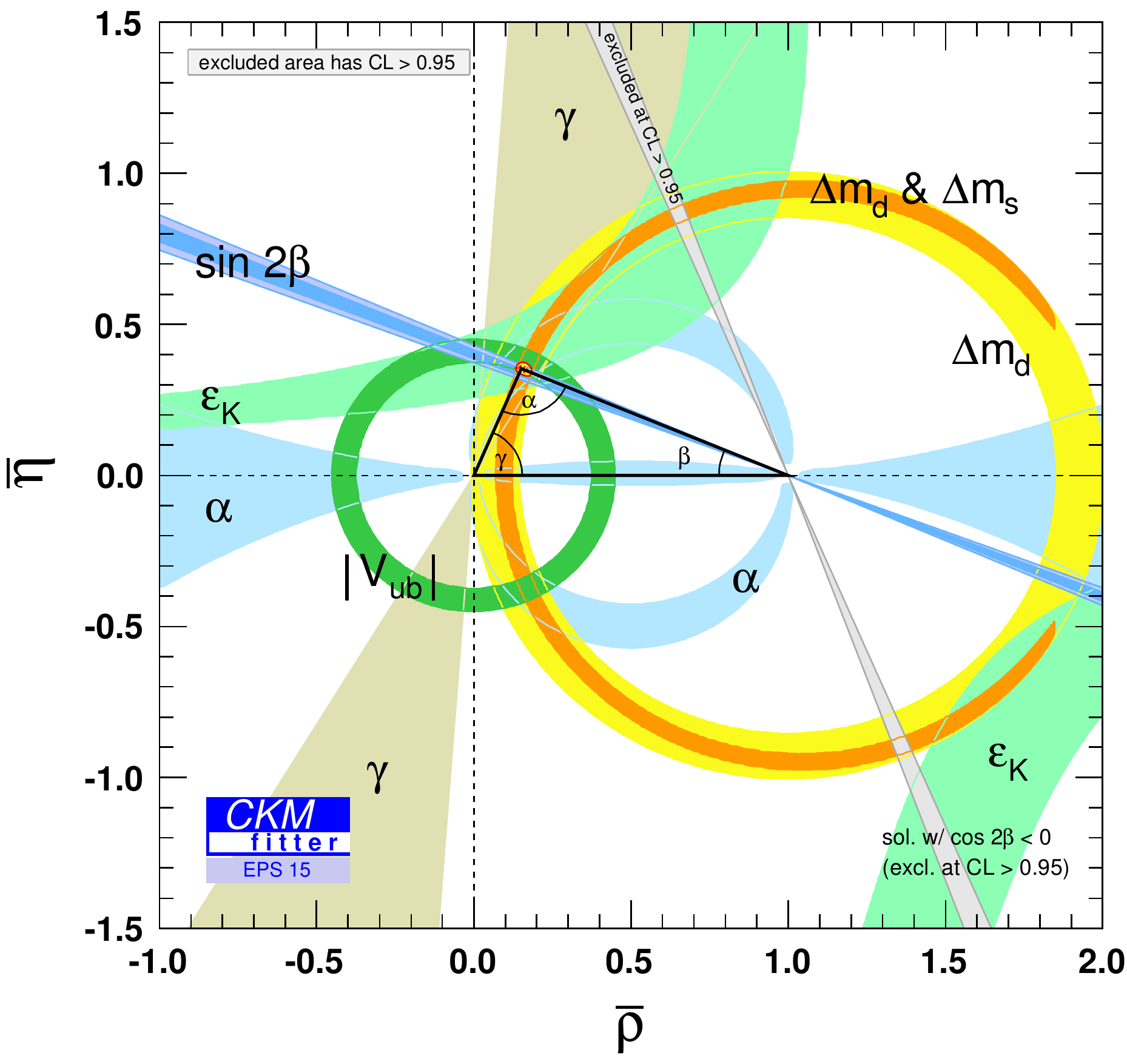}}
\caption{The standard model CKM fit, and individual constraints (colored regions
show 95\%~CL)~\cite{Hocker:2001xe}.}
\label{fig:SMCKMfit}
\end{figure}

A detailed introduction to flavor physics is omitted here, as well as a review
of the determinations of CKM elements; see, e.g., Refs.~\cite{Ligeti:2015kwa,
PDG}. The magnitudes of CKM elements are mainly extracted from semileptonic and
leptonic $K$, $D$, and $B$ decays, and $B_{d,s}$ mixing.  These determine the
sides of the unitarity triangle shown in Fig.~\ref{fig:SMCKMfit}, which is a
convenient way to compare many constraints on the SM and visualize the level of
consistency.  Any constraint which renders the area of the unitarity triangle
nonzero, such as nonzero angles (mod $\pi$), has to measure $CP$ violation, and
were reviewed in another talk~\cite{KMtalk}.  Some of the most important
measurements are shown in Fig.~\ref{fig:SMCKMfit}, together with the CKM fit in
the SM.  (The notation $\rhobar,\, \etabar$ instead of $\rho,\, \eta$ simply
corresponds to a small modification of the original Wolfenstein parametrization,
to keep unitarity exact.)  While Fig.~\ref{fig:SMCKMfit} shows very good
consistency, it does not address how large new physics contributions are
allowed.  As we see below, in the presence of new physics the fit becomes less
constrained, and ${\cal O}(20\%)$ NP contributions to most FCNC processes,
relative to the SM, are still allowed.

\begin{figure}[t]
\centerline{\includegraphics[width=.5\textwidth]{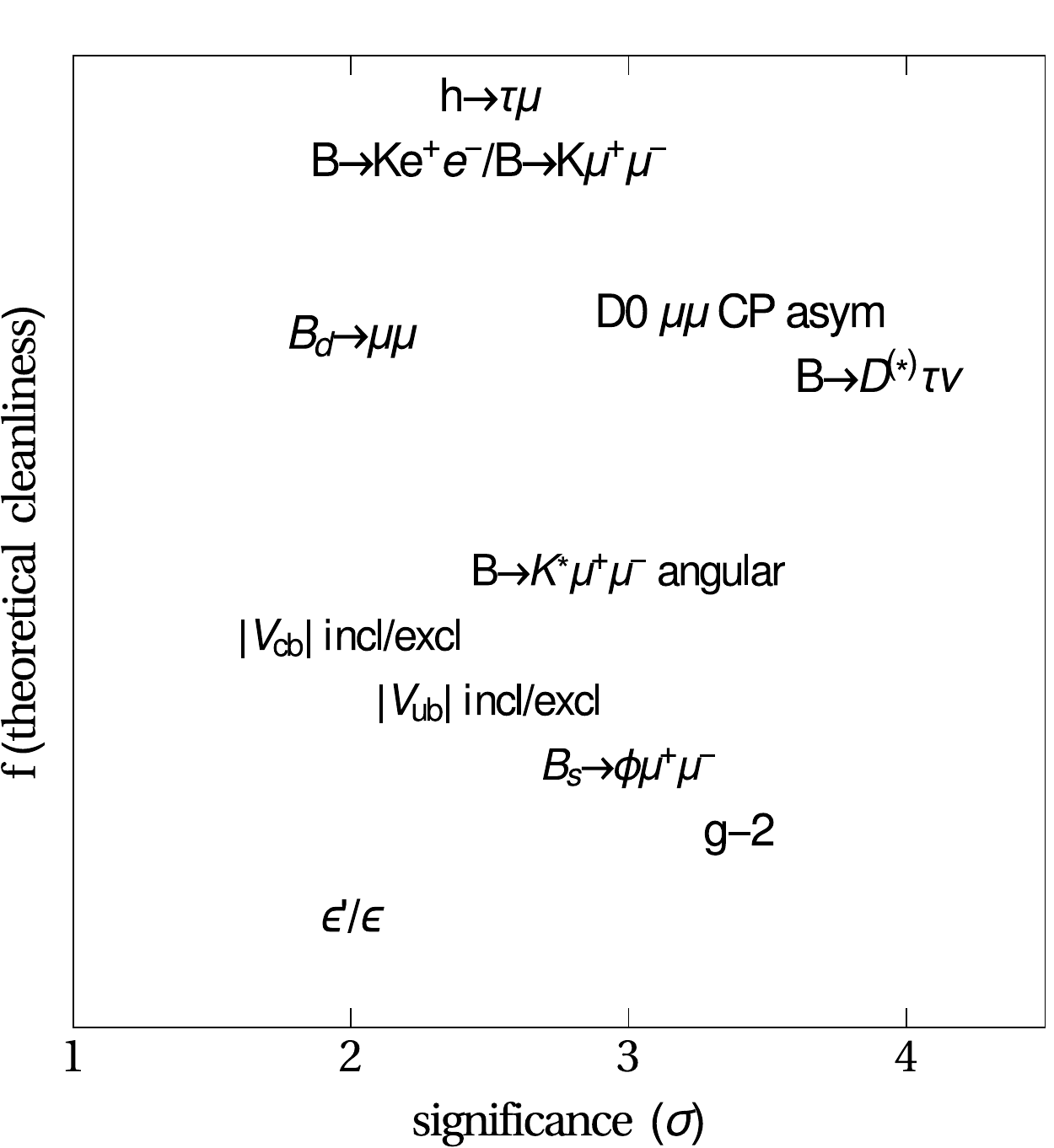}}
\caption{Some recent measurements in tension with the SM.  The horizontal axis
shows the nominal significance.  The vertical axis shows (monotonically, in my
opinion) an undefined function of an ill-defined variable: the theoretical
cleanliness.  That is, the level of plausibility that a really conservative
estimate of the theory uncertainty of each observable may affect the
significance of its deviation from the SM by $1\sigma$.}
\label{fig:cartoon}
\end{figure}

Several measurements show intriguing deviations from the SM predictions.  Some
of those that reach the $2-4\,\sigma$ level are depicted schematically in
Fig.~\ref{fig:cartoon}.  The horizontal axis shows the nominal significance and
the vertical axis relates to the theoretical cleanliness of the SM predictions. 
What I mean is some (monotonic) measure of the plausibility that a conservative
estimate of the theory uncertainty may affect the overall significance by
$1\sigma$.  All of these are frequently discussed, some have triggered hundreds
of papers, and could be the subjects of entire talks each.

\begin{figure}[tb]
\centerline{\scriptsize
\begin{tabular}{c|cc}
\hline \hline
&  $R(D)$  &  $R(D^*)$ \\
\hline
BaBar  &  $0.440 \pm 0.058 \pm 0.042$  &  $0.332 \pm 0.024 \pm 0.018$ \\
Belle  & $0.375 \pm 0.064 \pm 0.026$  &  $0.293 \pm 0.038 \pm 0.015$ \\
Belle  &    &  $0.302 \pm 0.030 \pm 0.011$ \\
LHCb  &  &  $0.336 \pm 0.027 \pm 0.030$  \\
\hline
Exp. average  &  $0.397 \pm 0.040 \pm 0.028$  &  $0.316 \pm 0.016 \pm 0.010$ \\
\hline
my SM expectation  &  $0.300 \pm 0.010$  &  $0.252 \pm 0.005$  \\
\hline
Belle~II, 50/ab  &  $\pm 0.010$  &  $\pm 0.005$\\
\hline\hline
\end{tabular}\hfill
\raisebox{-64pt}{\includegraphics[width=.42\textwidth, clip, 
  bb=0 5 555 375]{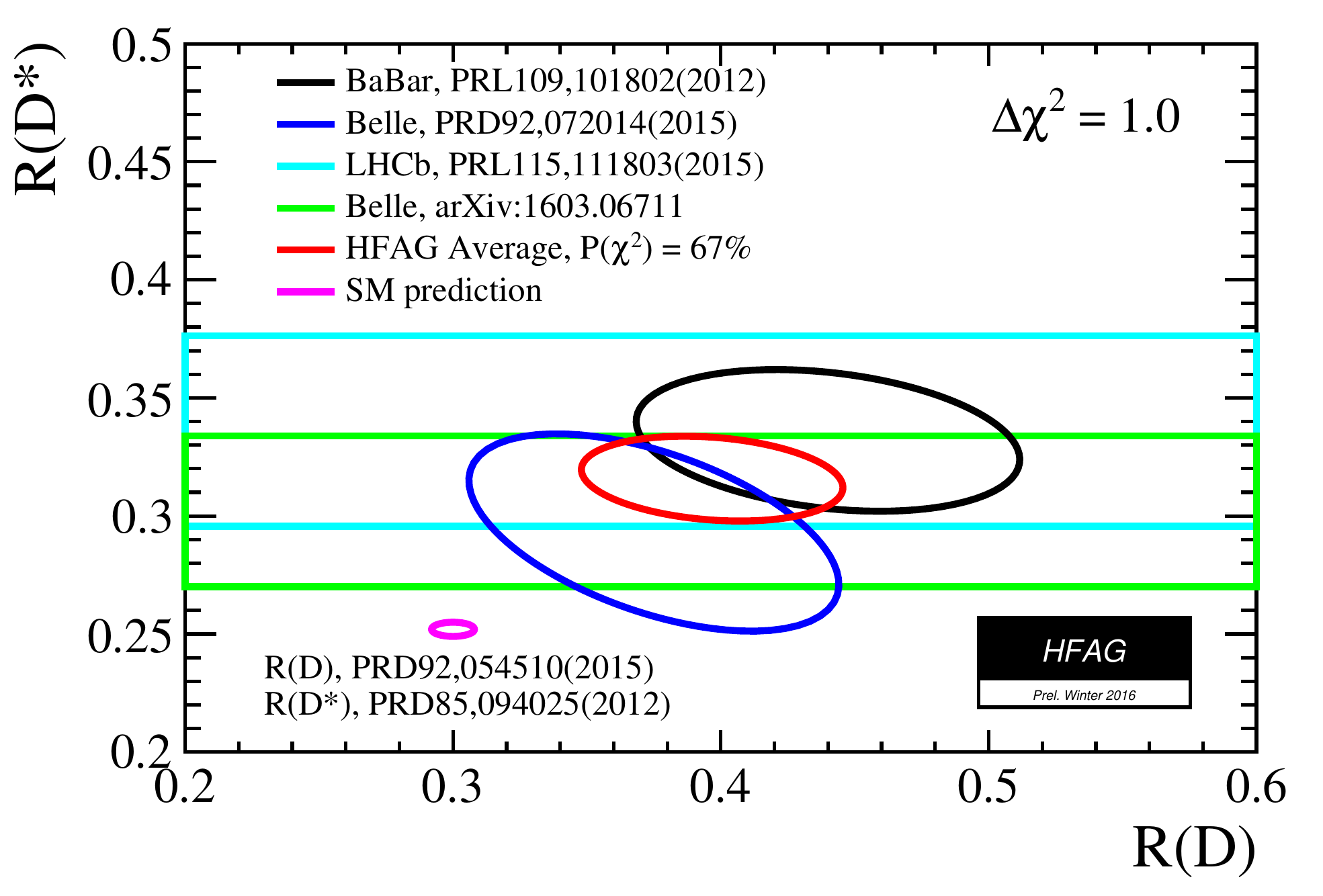}}}
\caption{Left: measurements of $R(D^{(*)})$~\cite{Lees:2012xj, Huschle:2015rga,
Aaij:2015yra, Abdesselam:2016cgx}, their averages~\cite{HFAG}, the SM
predictions~\cite{Lattice:2015rga, Na:2015kha, Fajfer:2012vx}, and future
sensitivity~\cite{Belle2predictions}.  Right: the measurements, world
average (red), and SM prediction (magenta).}
\label{fig:RDdata}
\end{figure}

Currently, the $B\to D^{(*)}\tau\bar\nu$ rates, specifically the $R(D^{(*)}) =
\Gamma(B\to D^{(*)}\tau\bar\nu) / \Gamma(B\to D^{(*)}l \bar\nu)$ ratios (where
$l=e,\, \mu$) constitute the most significant discrepancy from the SM in
collider experiments~\cite{Lees:2012xj, Lees:2013udz, Huschle:2015rga,
Aaij:2015yra, Abdesselam:2016cgx} (aside from neutrino masses).  The effect is
at the $4\sigma$ level~\cite{HFAG}.  Figure~\ref{fig:RDdata} shows the current
data, the SM expectations, as well as the expected Belle~II sensitivity.  These
measurements show good consistency with one another.  The theory is also on
solid footing, since heavy quark symmetry suppresses model independently the
hadronic physics needed for the SM prediction, most of which is actually
constrained by the measured $B\to D^{(*)}l \bar\nu$ decay distributions.

It is somewhat surprising to find such large deviations from the SM in processes
which occur at tree level in the SM.  The central values of the current world
averages would imply that there has to be new physics at a fairly low scale. 
Some scenarios are excluded by LHC Run 1 bounds already, and many more will soon
be constrained by LHC Run 2 data.  To fit the current central values, mediators
with leptoquark or $W'$ quantum numbers are preferred compared to scalars. 
Leptoquarks are favored if one requires the NP to be minimally flavor violating
(MFV), which helps explain the absence of other flavor signals and suppress
direct production of the new particles at the LHC from partons abundant in
protons~\cite{Freytsis:2015qca}.  There are several options for the
lepton-flavor structure of the new physics, which can have ``lepton-MFV" or 
``$\tau$-alignment"~\cite{Freytsis:2015qca}.  For example, the latter can be
realized with an $A_4$-type symmetry for the leptons, which links it to neutrino
flavor~\cite{Varzielas:2015iva}.  To illustrate the wide range of possibilities,
there are viable scenarios in which $B\to D^{(*)}\tau\bar\nu$ are SM-like, but
$B\to D^{(*)} l \bar\nu$ are suppressed by interference between NP and the
SM~\cite{future}.

There are many further experimental measurements that can be done to clarify
this anomaly.  The $B\to D^{(*)}\tau\bar\nu$ rates seem to
exceed~\cite{Freytsis:2015qca} the LEP measurements of the inclusive $b\to X
\tau \bar\nu$ rate~\cite{PDG}, and the inclusive $B\to X_c\tau\nu$
rate~\cite{Ligeti:2014kia} has not yet been measured.  The equality of the $e$
and $\mu$ rates are not well constrained, and the currently allowed
differences~\cite{Aubert:2008yv, Dungel:2010uk} open up (or keep open) model
building options~\cite{Greljo:2015mma}.  In many scenarios, bounds on $b\to
s\nu\bar\nu$ processes are very important~\cite{Freytsis:2015qca,
Fortes:2015jaa}.  A lot will be learned, hopefully soon, from LHCb result on
$R(D)$, measurements using hadronic $\tau$ decays, measurements in $\Lambda_b$
and $B_s$ decays, and later from Belle~II.  If a deviation from the SM is
established, it will strongly motivate to measure all possible semitauonic
modes, both in $b\to c$ and $b\to u$ transitions~\cite{Bernlochner:2015mya,
Hamer:2015jsa}.

\begin{figure}[t]
\centerline{\includegraphics[width=.5\textwidth]{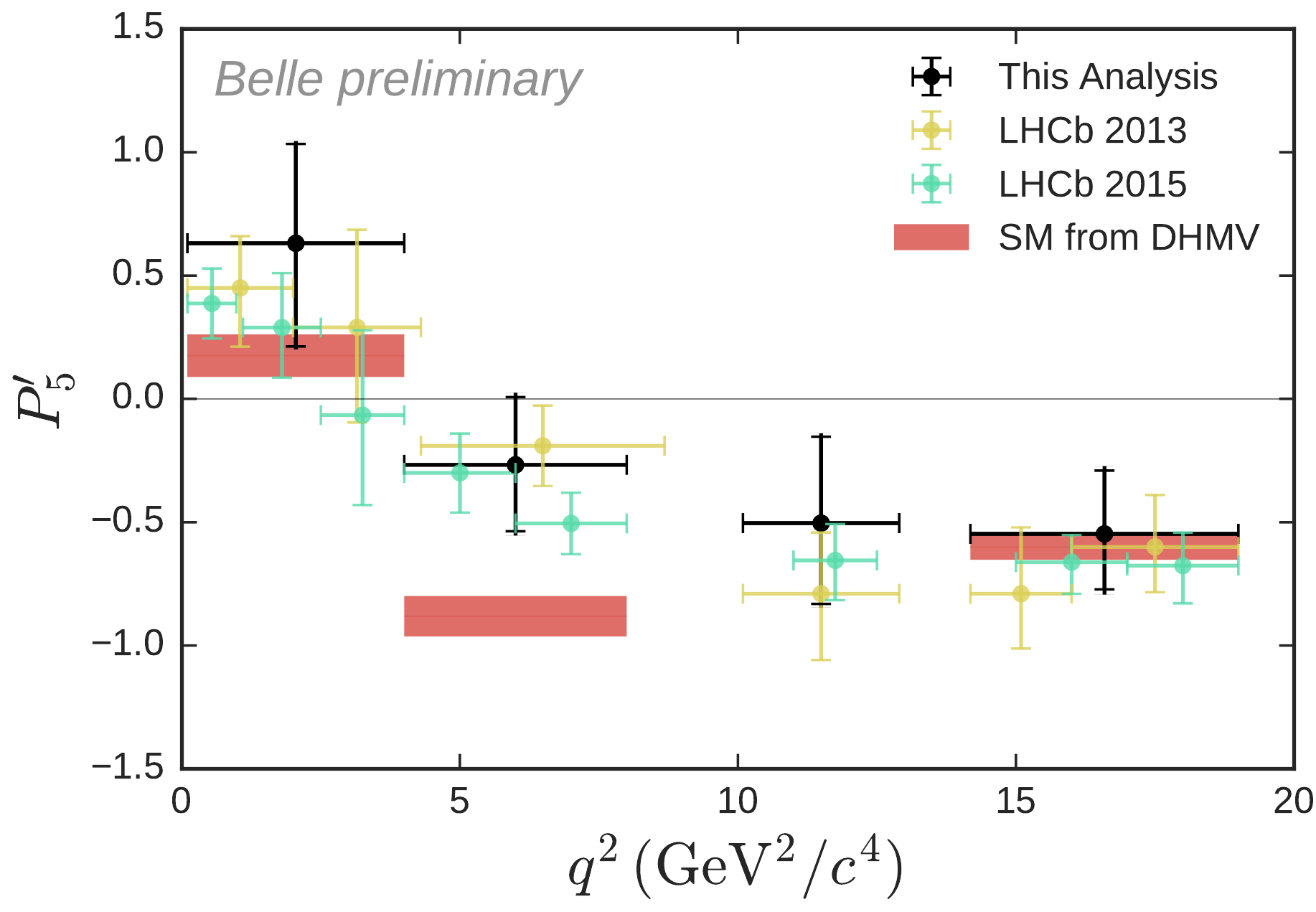}\hfill
\includegraphics[width=.47\textwidth]{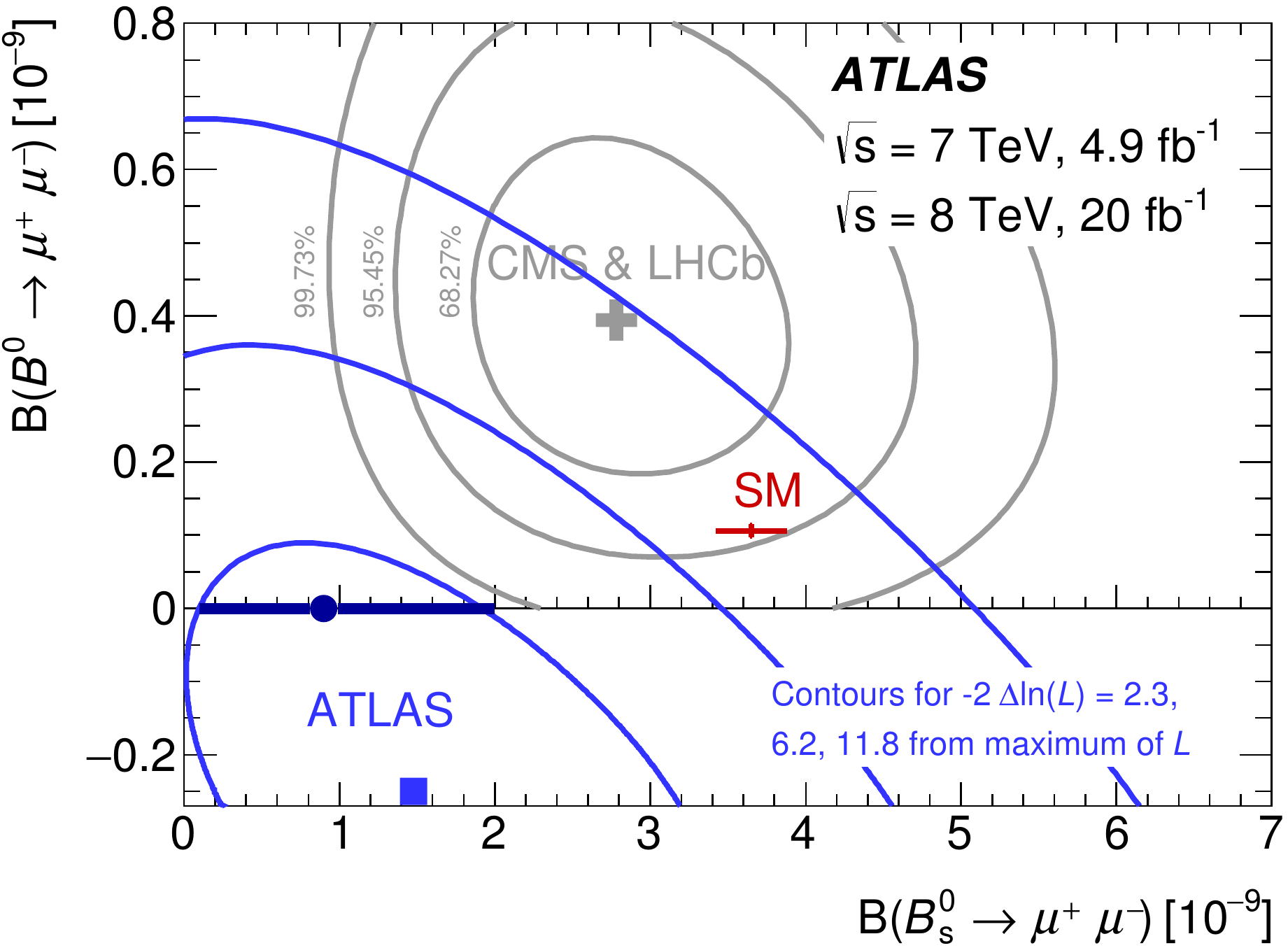}}
\caption{Left: The LHCb~\cite{Aaij:2015oid} and Belle~\cite{Abdesselam:2016llu}
measurements of $P'_5$ in $B\to K^*\mu^+\mu^-$. 
Right: The $B_{s,d} \to\mu^+\mu^-$ measurements from LHCb and
CMS~\cite{CMS:2014xfa}, and the ATLAS constraints~\cite{Aaboud:2016ire}
superimposed.}
\label{fig:KsBsmumu}
\end{figure}

Another measurement which has drawn immense attention is the ``$P'_5$ anomaly"
in a $B\to K^*\mu^+\mu^-$ angular distribution (see, e.g.,
Refs.~\cite{Descotes-Genon:2013wba, Altmannshofer:2014rta}), measured at
LHCb~\cite{Aaij:2015oid} and recently at Belle~\cite{Abdesselam:2016llu}, and
discussed in another talk in more detail~\cite{GLtalk}.  The measurements are
shown in the left plot in Fig.~\ref{fig:KsBsmumu}, together with a SM
prediction~\cite{Descotes-Genon:2014uoa}.  These ``optimized observables" are
based on the SCET factorization theorem for semileptonic $B$ decay form
factors~\cite{Bauer:2002aj, Beneke:2003pa}, and constructing combinations from
which the ``nonfactorizable" (``soft") contributions cancel.  (These are
nonperturbative functions of $q^2$, which obey symmetry
relations~\cite{Charles:1998dr}; additional terms are either power suppressed or
contain an explicit $\alpha_s$ factor.)  The magnitudes of the correction terms,
that is one's ability to calculate the form factor ratios at small $q^2$
reliably, is debated~\cite{Jager:2014rwa} (and are not well constrained by data
yet).  The tension between theory and the data is certainly intriguing, and many
studies exist both in terms of model independent fits and specific model
predictions.  Some of the simplest models are $Z'$-like, with nonuniversal
flavor couplings.  One may be concerned that the best fit is a new contribution
to the operator $O_9 = e^2 (\bar s \gamma_\mu P_L b) (\bar\ell \gamma^\mu \ell)$
in the effective Hamiltonian, the same term which would be modified if
theoretical control over the $c\bar c$ loop contributions were worse than
expected.  (This was also emphasized recently in Ref.~\cite{Ciuchini:2015qxb}.) 
There are many possible connections to the $\sim 2.5\sigma$ anomaly in
$\Gamma(B\to K e^+e^-) \neq \Gamma(B\to K \mu^+\mu^-)$ as well~\cite{Gudrun}.

For these observables, too, I trust that with improved measurements and theory,
the source of the currently seen effects will be understood.  With more data,
one can test  the $q^2$ (in)dependence of the extracted Wilson coefficients. In
the large $q^2$ (small recoil) region one can make model independent predictions
both for exclusive~\cite{Bobeth:2012vn} inclusive~\cite{Ligeti:2007sn} $b\to s
l^+l^-$ mediated decays, which is complementary to the small $q^2$ region, and
has different theory uncertainties.

Another anomaly observed is the $B_s\to \phi\mu^+\mu^-$ rate in the $1 < q^2 <
6\, \GeV^2$ region being about $3\,\sigma$ below theoretical
calculations~\cite{Aaij:2015esa}.  This relies on QCD sum
rules~\cite{Straub:2015ica} combined with lattice QCD calculations of the form
factors at large $q^2$~\cite{Horgan:2015vla}.  Extending the lattice results to
lower $q^2$ would help clarify the picture, as well as more precise
measurements, also at high $q^2$.

If new physics is at play in these processes, it is likely to impact $B\to
\mu^+\mu^-$, too.  The combined LHCb and CMS observation~\cite{CMS:2014xfa} of
$B_s \to \mu^+\mu^-$, the constraint on $B_d \to \mu^+\mu^-$, and the recent
ATLAS~\cite{Aaboud:2016ire} constraints are shown in the right plot in
Fig.~\ref{fig:KsBsmumu}, as well as the SM prediction.  Measuring a rate at the
$3\times 10^{-9}$ level is impressive, and future refinements are high
priority.  The nonperturbative input in this case is just $f_B$, which is under
good control in lattice QCD.

\begin{figure}[t]
\centerline{\includegraphics[width=.48\textwidth]{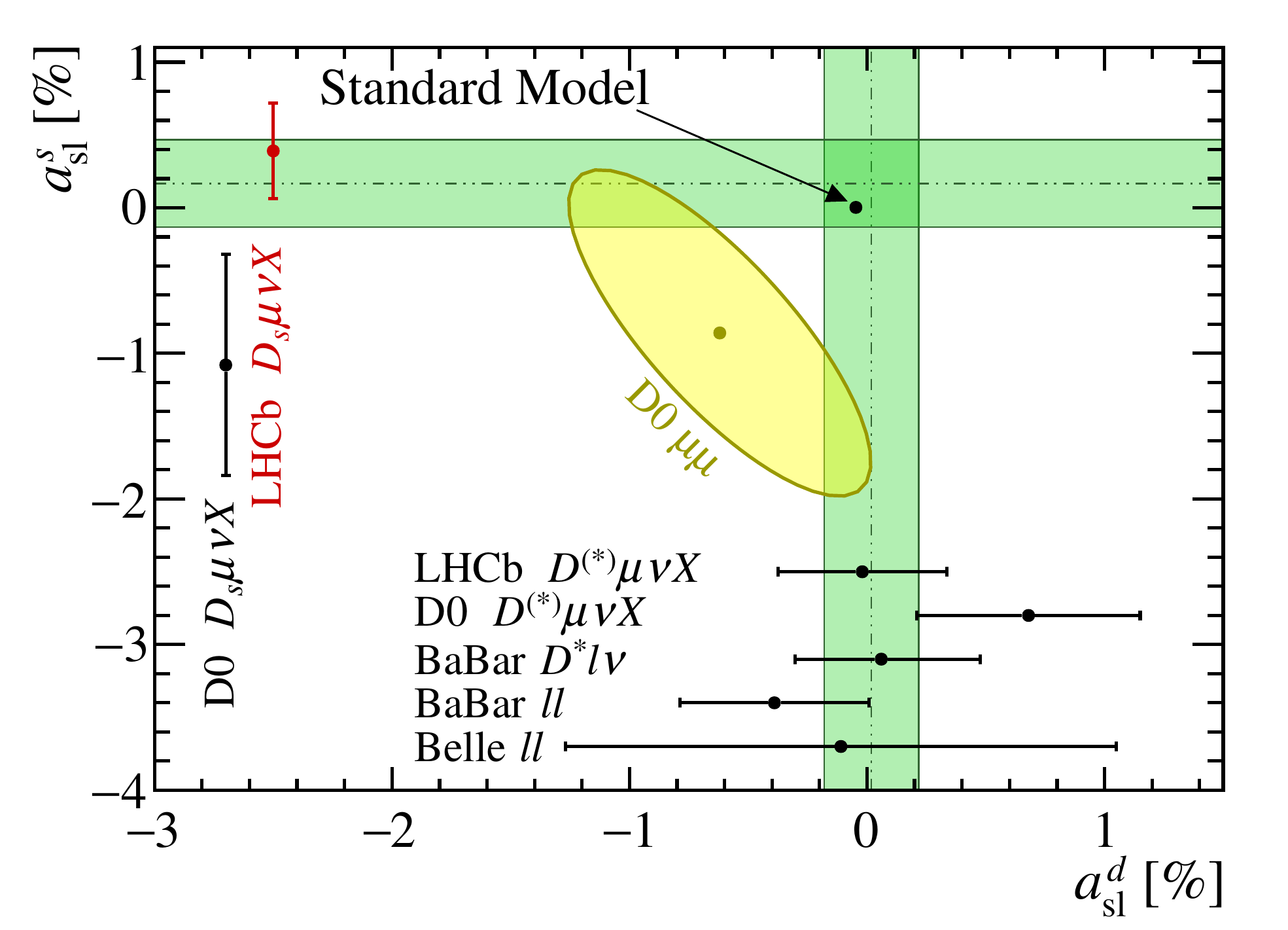} \hfil\hfil
\raisebox{4pt}{\includegraphics[width=.5\textwidth]{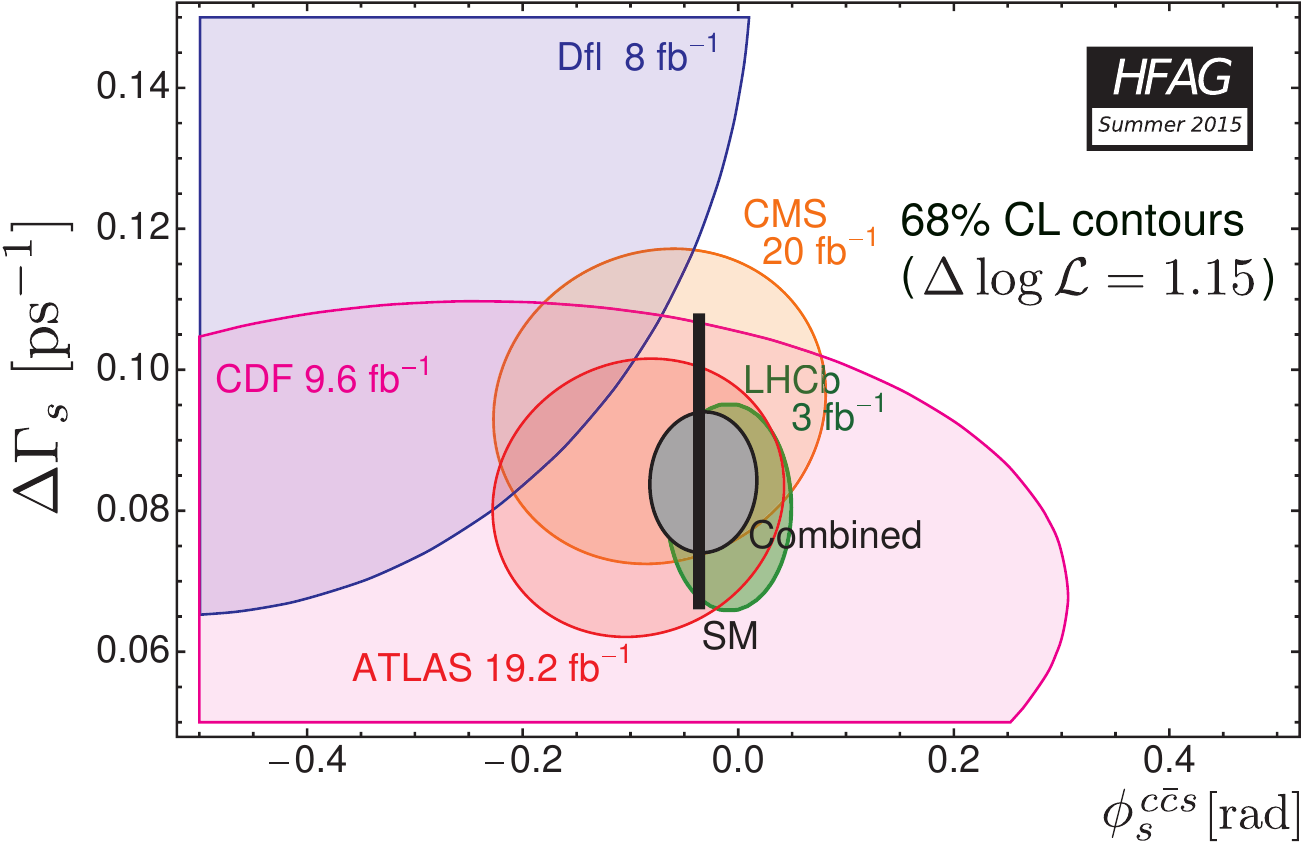}}}
\caption{Left: bounds on $CP$ violation in $B_{d,s}$ mixing, $a_{\rm
SL}^{d,s}$~\cite{Aaij:2016yze}. The vertical and horizontal bands show the
averages of the separate $B_d$ and $B_s$ measurements, respectively, and the
yellow ellipse is the D\O\ measurement.   Right: measurements of $\phi_s \equiv
-2\beta_s$ showing good consistency with the SM.}
\label{fig:cpvmix}
\end{figure}

Another deviation from the SM expectations, which is theoretically very clean,
and has been $3-4\,\sigma$, is the D\O\ measurement of the like-sign dimuon
charge asymmetry in semileptonic decays of $b$ hadrons, $(N_{\mu^+\mu^+} -
N_{\mu^-\mu^-}) / (N_{\mu^+\mu^+} + N_{\mu^-\mu^-})$~\cite{Abazov:2011yk}, shown
in the left plot in Fig.~\ref{fig:cpvmix}.  A nonzero signal could come from a
linear combination of $CP$ violation in $B_s$ and $B_d$ mixing, $a_{\rm
SL}^{d,s}$ (see, e.g., Ref.~\cite{Ligeti:2010ia}), and the SM prediction is well
below the current sensitivity.  Separate measurements of $a_{\rm SL}^d$ and
$a_{\rm SL}^s$ from BaBar, Belle, D\O, and LHCb are consistent in with the SM. 
The very recent LHCb measurement of $a_{\rm SL}^s = (0.39 \pm 0.33)\%$ with
3/fb~\cite{Aaij:2016yze}, reducing the uncertainty from 0.62\% with 1/fb, starts
to be in tension with the D\O\ anomaly.  If there is new physics in $CP$
violation in $B_s$ mixing, then one may also expect to see a deviation from the
SM in the time-dependent $CP$ asymmetry in $B_s\to \Jpsi\phi$.  Recent LHC
measurements, however, are consistent with the SM, as shown in the right plot in
Fig.~\ref{fig:cpvmix}.  Most importantly, the theory uncertainties are well
below the experimental sensitivity in the coming years, so a lot can be learned
from more precise measurements.

Understanding the long-standing tensions between inclusive and exclusive
measurements of $|V_{cb}|$ and $|V_{ub}|$ are important for NP searches. The
ratio $|V_{ub} / V_{cb}|$ together with $\gamma$ determine the apex of the
unitarity triangle from tree-level processes, which is crucial for improving the
sensitivity to NP in $B$ mixing and in $CP$ violation measurements involving
loop processes.  Understanding the QCD dynamics of semileptonic $B$ decays is
also important, because the theoretical tools coincide with those used in
inclusive and exclusive $b\to s\gamma$ and $b\to s \ell\bar\ell$ decays.  While
I have also entertained modifications of these CKM measurements due to NP (such
as right-handed currents~\cite{Bernlochner:2014ova}), many known theoretical and
experimental improvements can take place in the future, such as doing all
measurements in events where the other $B$ is fully reconstructed.

The muon $g-2$ measurement has a more than $3\sigma$ tension with the SM,
according to most analysis.  It may be a sign of new physics, but some
complicated strong interaction dynamics could still be at play and decrease the
significance of the deviation.  One hopes that lattice QCD will determine
reliably the hadronic light-by-light scattering and vacuum polarization
contributions.  While supersymmetric models with relatively light sleptons can
still account for the deviation~\cite{deVries:2015hva}, the LHC will improve
limits on such NP explanations, and sub-GeV weakly coupled new particles could
also be at play~\cite{MPtalk}.  For the heavy NP explanation, somewhat puzzling
is that the required new physics contribution is the same size as the one-loop
SM electroweak contribution.

It has long been known that kaon $CP$ violation is sensitive to some of the
highest energy scales.  For the $\epsilon$ parameter, the SM is in good
agreement with the data, and the NP contribution is constrained to be $\,\lsim
30\,\%$ of that of the SM~\cite{Ligeti:2016qpi}.  Calculating the SM prediction
for direct $CP$ violation, the $\epsilon'$ parameter, has been a multi-decade
challenge, and important progress has been made recently~\cite{Bai:2015nea}. 
Results with several lattice spacings are needed to decide if NP is present.

These experimental hints of possible deviations from the SM are fantastic for
several reasons.  Unambiguous evidence for NP would obviously be the start of a
new era, and one would also get a rough upper bound on the scale of NP, even if
it is not seen directly at ATLAS \& CMS.  It is also useful to have experimental
results challenge theory, since unexpected signals motivate both model building
and revisiting the SM predictions.  This was the case with the Tevatron anomaly
in the $t\bar t$ forward-backward asymmetry, $A_{\rm FB}^{t\bar t}$, which
disappeared due to refinements of the experimental results (the SM predictions
also improved~\cite{Czakon:2014xsa}).  Concerning the recent $3\,\sigma$ hint
for direct $CP$ violation in the difference of $CP$ asymmetries in $D\to K^+K^-$
and $D\to \pi^+\pi^-$, $\Delta A_{CP} = A_{K^+K^-} - A_{\pi^+\pi^-}$, I doubt
the initial measurement near $1\%$ could be attributed to the
SM~\cite{Isidori:2011qw}.  The central value of the world average has decreased
since 2012, as has the significance of the hint for $\Delta A_{CP} \neq 0$.  We
probably still do not know how large $\Delta A_{CP}$ the SM could generate. 
However, exploring it taught us, for example, about how much (or how little) the
quark and squark mixing matrices can differ and squark masses (don't) need to
be degenerate~\cite{Gedalia:2012pi, Mahbubani:2012qq} in alignment
models~\cite{Nir:1993mx}.

\begin{figure}[t]
\centerline{\includegraphics[width=.52\textwidth]{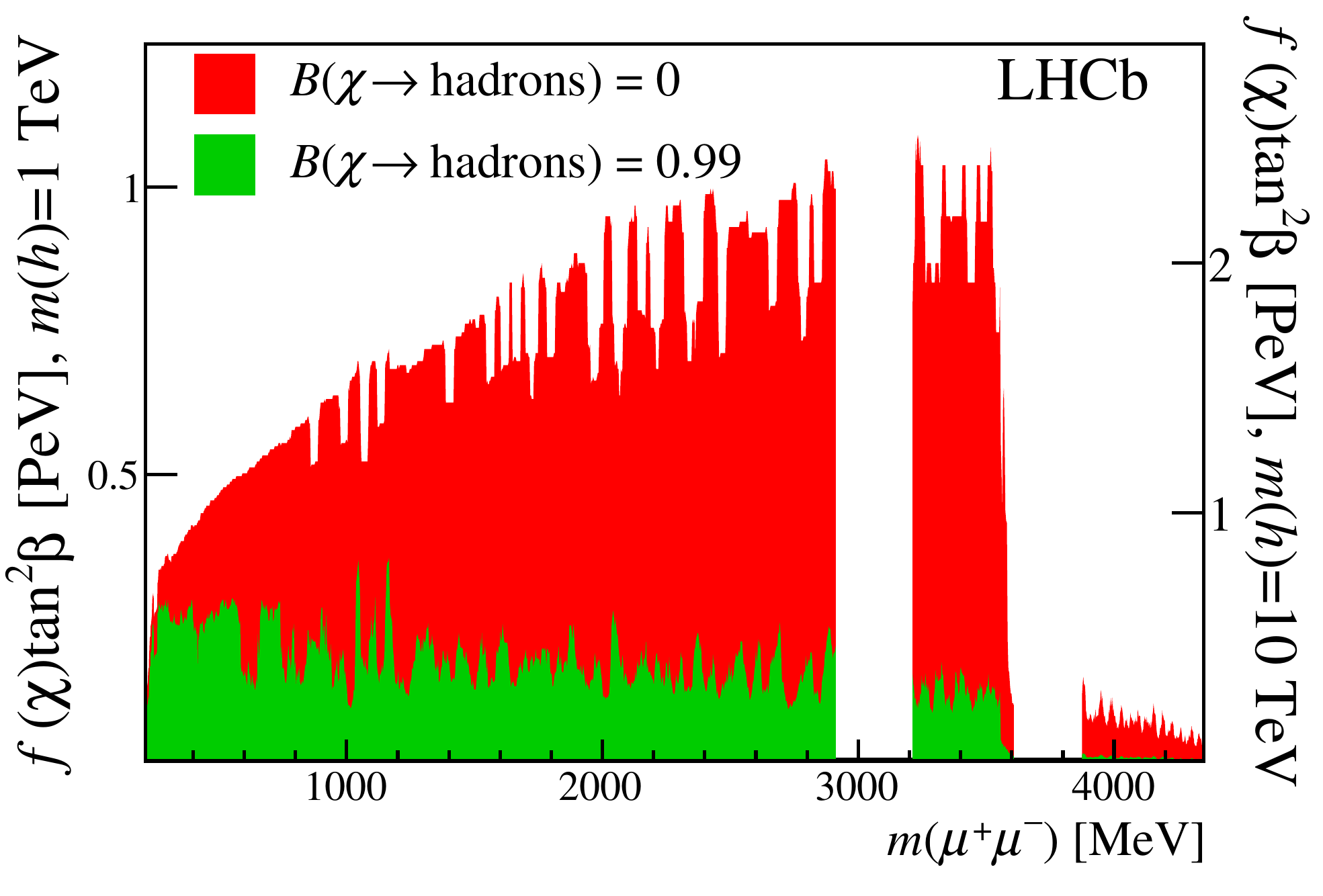} \hfill
\includegraphics[width=.45\textwidth]{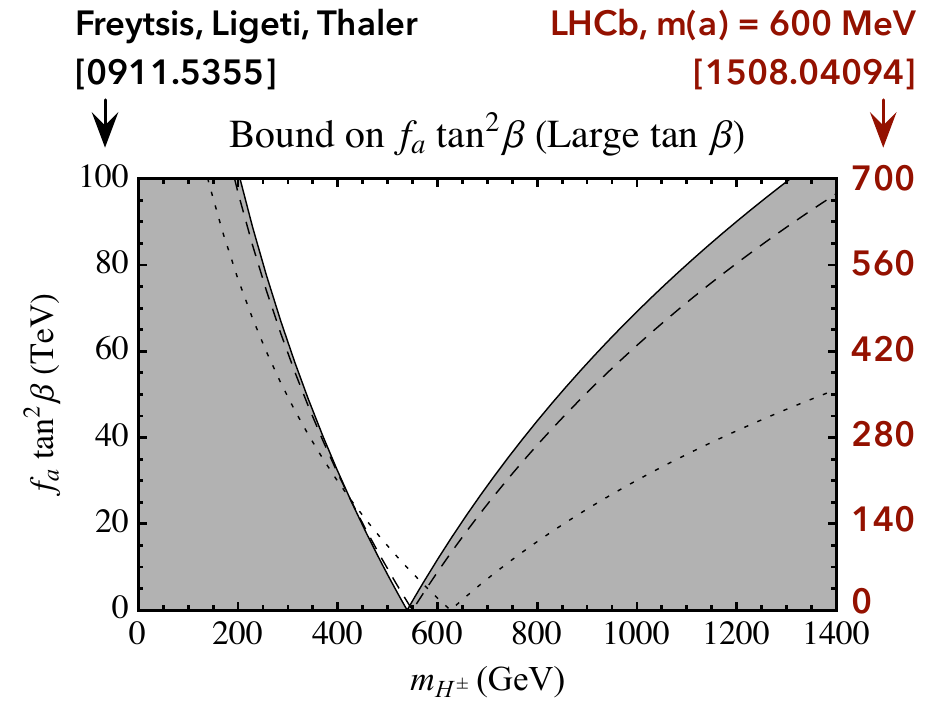}}
\caption{Left: LHCb bounds on $f_\chi^2\, \tan^2\beta$ as a function of
$m_{\mu+\mu^-}$~\cite{Aaij:2015tna} in the model~\cite{Freytsis:2009ct}.
Right: The bound as a function of $m_{H^\pm}$ in the same model; the right axis
shows a nearly order of magnitude improvement.}
\label{fig:axionsearch}
\end{figure}

A recent measurement in which no anomaly is seen, but I find the nearly order of
magnitude increase in mass-scale sensitivity due to a recent LHCb
analysis~\cite{Aaij:2015tna} impressive, is the search for an axion-like
particle, coupling to SM fermions as $(m_\psi/f_a)\, \bar\psi\gamma_5\psi\, a$,
explained in another talk~\cite{GLtalk}.  Such models are also interesting,
because they may have highly suppressed spin-independent direct detection cross
sections~\cite{Freytsis:2010ne}. The left plot in Fig.~\ref{fig:axionsearch}
shows the 95\% CL lower bound on $f_\chi^2\, \tan^2\beta$ in the model of
Ref.~\cite{Freytsis:2009ct}, from the absence of a narrow $\mu^+\mu^-$ peak in
$B\to K^*\chi\,(\chi\to \mu^+\mu^-)$ as a function of $m_{\mu^+\mu^-}$.  The
bound is shown for $m_{H^\pm} = 1\,\TeV$ and two values of the hadronic
branching fraction of the axion-like particle.  The right plot shows the bound
on the same quantity as a function of $m_{H^\pm}$ ($f_a$
in~\cite{Freytsis:2009ct} is $f_\chi$ in~\cite{Aaij:2015tna}).  The left
vertical axis is the bound estimated in 2009~\cite{Freytsis:2009ct} from BaBar
\& Belle data with only a few bins, and the right vertical axis shows the LHCb
bound~\cite{Aaij:2015tna}.  The dashed (dotted) curve shows the bound for
$\tan\beta = 3$ ($\tan\beta = 1$).  In this model, for any value of $\tan\beta$,
the NP contribution vanishes due to a cancellation for a certain
value~of $m_{H^\pm}$.

\section{Future increases in new physics scales probed}
\label{sec:future}

I would like to talk about three topics briefly in this part: (i) the future
theory uncertainty of the measurement on $\sin2\beta$ from $B\to \Jpsi K_S$;
(ii) the future sensitivity to NP in mixing of neutral mesons; (iii) sensitivity
of flavor physics experiments to very heavy vector-like fermions.

\subsection{What are the ultimate theory uncertainties of
$\mathbf{\sin2\beta}$?}

The theoretical uncertainty of the SM predictions for the time dependent $CP$
asymmetries in the ``gold-plated" modes $B\to \Jpsi K_S$ and $B_s\to \Jpsi \phi$
are of great importance.  They arise from contributions to the decay amplitude
proportional to $V_{ub}V_{us}^*$ instead of the dominant $V_{cb}V_{cs}^*$ terms.
I refer to this as $V_{ub}$ contamination, instead of the often used penguin
pollution phrase (which is less correct and less clear).  This effect did not
matter in practice in the past, but it will be important for interpreting the
full LHCb and Belle II data sets.   During the BaBar/Belle era, the experimental
precision was an order of magnitude above the nominal magnitude of the
theoretical uncertainty, $\lambda^2 (\alpha_s/\pi) \sim 0.004$.  So even a
factor of few enhancement of the latter did not matter.\footnote{Until 1997 this
estimate was often written as $\lambda^2 (\alpha_s/4\pi)$. Omitting the factor 4
anticipates some enhancement of the penguin matrix element, observed in
charmless $B$ decays~\cite{Godang:1997we} but not yet well constrained in decays
to charmonia. Better calculable ${\cal O}(10^{-3})$ effects arise from $CP$
violation in $K$ and $B$ mixing, and the $\Gamma_{B_L} - \Gamma_{B_H}$ width
difference~\cite{Grossman:2002bu}.}  A number of approaches have been developed,
using a combination of diagrammatic and flavor symmetry arguments with various
assumptions~\cite{Jung:2012mp, Frings:2015eva}.  (I hasten to add a triviality:
there is no relation based only on $SU(3)$ flavor symmetry between final states
which are entirely in different representations; e.g., $\phi$ is an $SU(3)$
singlet and $\rho$ \& $K^*$ are members of an octet.)  The experimental tests
performed so far~\cite{KMtalk} do not indicate big enhancements of the theory
uncertainties.

The question that really matters in my opinion is not what it takes to set
plausible upper bounds on the $V_{ub}$ contamination, while the measurements
agree with the SM, but what it would take to convince the community that NP is
observed at LHCb and Belle~II, especially if no NP is seen by ATLAS and CMS. 
Therefore, one cannot overemphasize the importance of starting from rigorous
theoretical foundations, with well defined expansion parameter(s).  

\begin{figure}[t]
\centerline{\includegraphics[width=.45\textwidth]{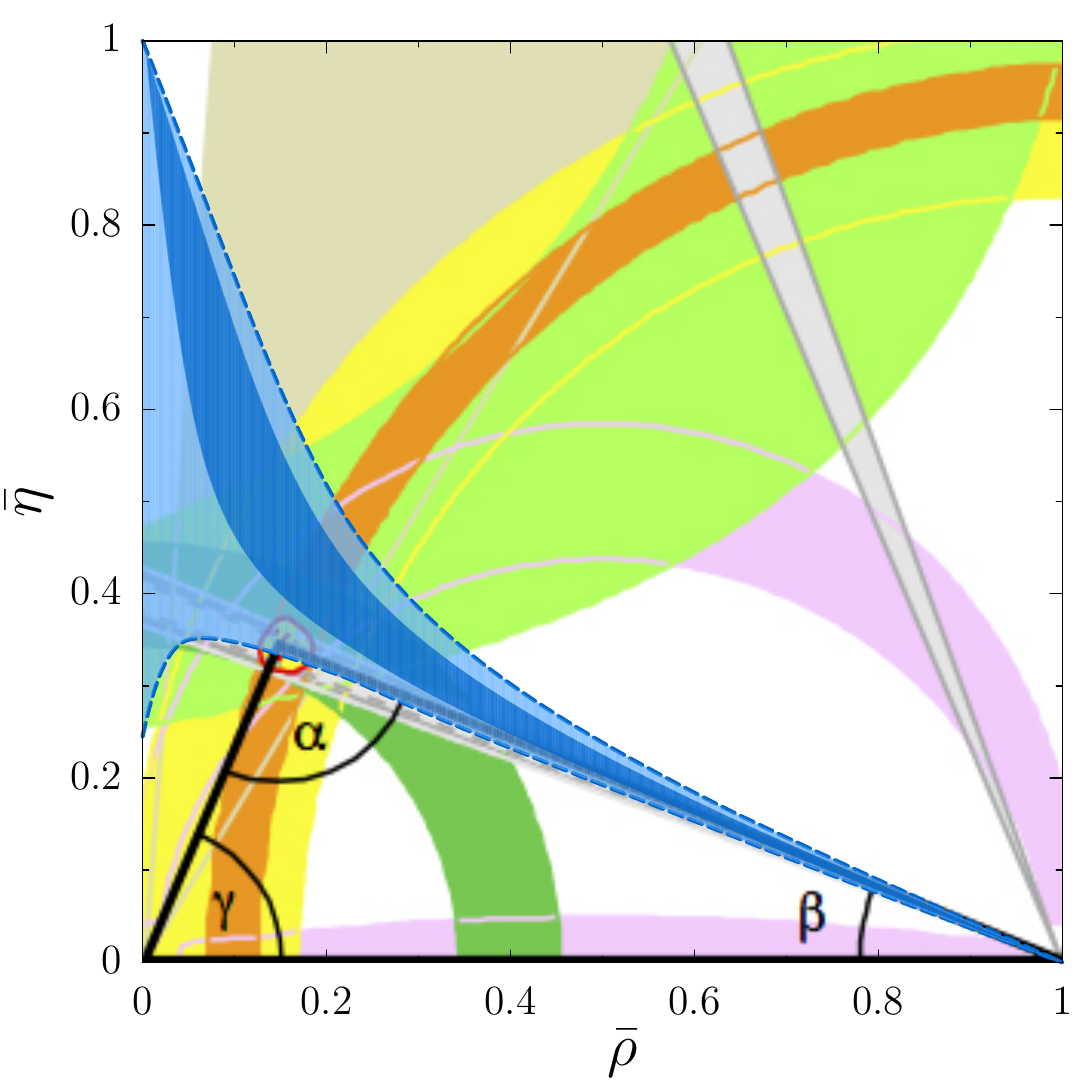}}
\caption{The dark (light) blue region shows the $1\sigma$ ($2\sigma$) constraint
in the $\rhobar - \etabar$ plane from
Eq.~(\protect\ref{beauty})~\cite{Ligeti:2015yma}.}
\label{fig:s2bfuture}
\end{figure}

A relation based only on $SU(3)$ flavor symmetry, which cancels the $V_{ub}$
contamination in $\sin2\beta$ against other observables in the $SU(3)$ limit,
was constructed recently~\cite{Ligeti:2015yma}
\beq\label{beauty}
\sin 2\beta = \frac{S_{K_S} - \lambda^2 S_{\pi^0}
  - 2(\Delta_K + \lambda^2 \Delta_\pi) \tan\gamma\, \cos2\beta}{1 +
  \lambda^2}\,.
\eeq
Here $S_h$ ($h=K,\,\pi$) is the usual coefficient of the $\sin(\Delta m\, t)$
term in the time-dependent $CP$ asymmetry~\cite{PDG} in $B\to \Jpsi\, h^0$,
$\lambda\simeq 0.225$ is the Wolfenstein parameter,
\beq
\Delta_h = \frac{\Gam{B_d}{h^0} - \Gam{B^+}{h^+}}
  {\Gam{B_d}{h^0} + \Gam{B^+}{h^+}} \,,
\eeq
and $\bar\Gamma$ denotes the $CP$ averaged rates.  Using Eq.~(\ref{beauty}) it
is possible to replace the $V_{ub}$ contamination in the $\sin2\beta \simeq
S_{K_S}$ relation with isospin breaking, which could be smaller than the
possibly enhanced $V_{ub}$ contamination one wants to constrain.  It also
provides redundancy, replacing one theory uncertainty with a different one.  For
the $V_{cb}V_{cs}^*$ terms in the effective Hamiltonian, $\Delta_{K,\pi}$
violate isospin, but the $V_{ub}V_{us}^*$ terms generate nonzero $\Delta_h$ even
in the isospin limit.  The resulting constraint on the $\rhobar - \etabar$ plane
is shown in Fig.~\ref{fig:s2bfuture}~\cite{Ligeti:2007sn}.

Measuring all terms in Eq.~(\ref{beauty}) is not straightforward.  Many of the
current measurements of $\Delta_h$ and the production asymmetry of $B^+B^-$ vs.\
$B^0\B0bar$ in $\Upsilon(4S)$ decay, $f_{+-}/f_{00}$, are circular (the
measurements of either assume that the other asymmetry
vanishes)~\cite{Jung:2015yma}, so the slight tension in Fig.~\ref{fig:s2bfuture}
should be interpreted with caution.  To disentangle $\Delta_h$ from the
production asymmetry, more precise measurements of the latter are needed.  One
option may be to utilize that isospin violation in inclusive semileptonic decay
is suppressed by $\lqcd^2/m_b^2$~\cite{Jung:2015yma}.  (Similar suppression of
$SU(3)$ symmetry breaking in inclusive $B$ decays by $\lqcd^2/m_b^2$ is the
basis for a theoretically clean prediction for the ratio $\Gamma(B\to
X_s\ell^+\ell^-) / \Gamma(B\to X_u\ell \bar\nu)$ at large
$q^2$~\cite{Ligeti:2007sn}.)

It is an open question how well it will be possible to ultimately constrain
(convincingly) the size of $V_{ub}$ contamination in the measurements of
$\sin2\beta$ and $2\beta_s (\equiv -\phi_s)$.  Given that $SU(3)$ flavor
symmetry has been used to analyze $B$ decays for decades, and previously unknown
$SU(3)$ relations can be discovered in 2015, makes me optimistic that a lot more
progress can be achieved.

\subsection{New physics in SM loop processes}

\begin{figure}[t]
\centerline{
\includegraphics[height=6cm,clip,bb=15 15 550 520]{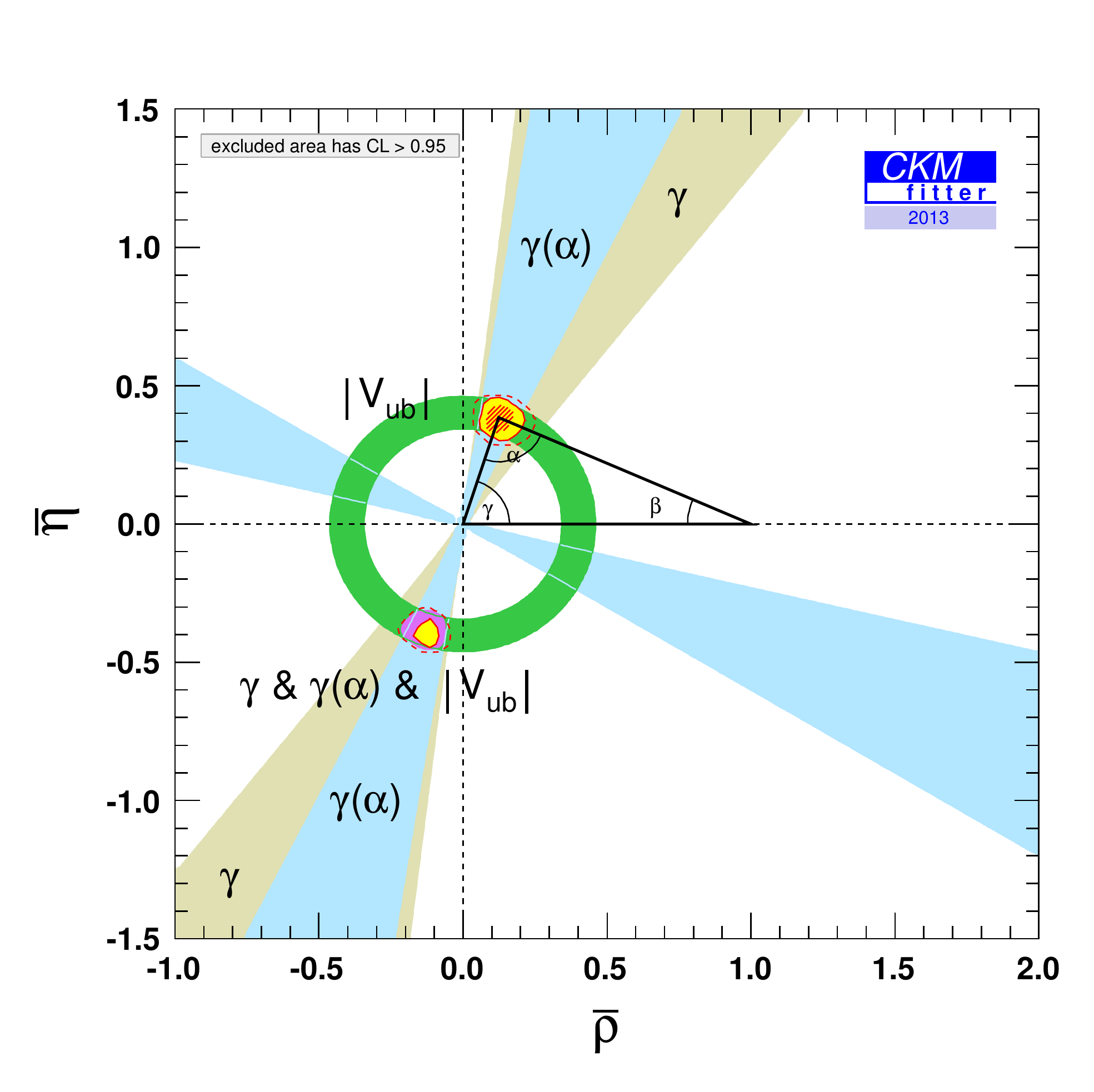}
\hfil\hfil
\includegraphics[height=6cm,clip,bb=105 20 670 550]{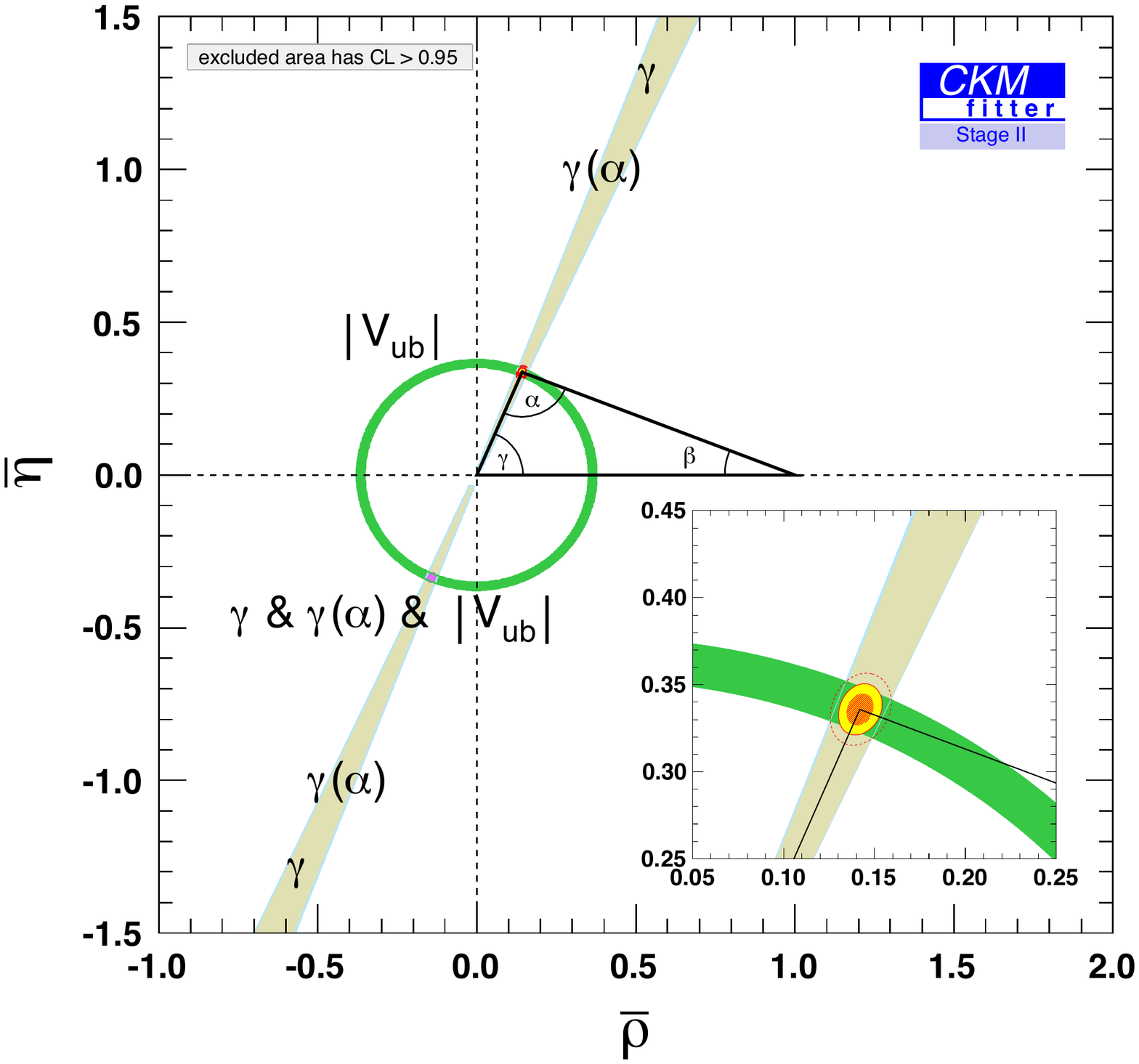}}
\caption{Constraints on $\rhobar - \etabar$, allowing NP in the $B_{d,s}$ mixing
amplitudes (left) and the expectation using 50\,ab$^{-1}$ Belle~II and
50\,fb$^{-1}$ LHCb data (right)~\cite{Charles:2013aka}.  Colored regions show
95\% CL, as in Fig.~\protect\ref{fig:SMCKMfit}.}
\label{fig:NPrhoeta}
\end{figure}

Although the SM CKM fit in Fig.~\ref{fig:SMCKMfit} shows impressive and
nontrivial consistency, the implications of the level of agreement are often
overstated.  Allowing new physics contributions, there are a larger number of
parameters related to $CP$ and flavor violation, and the fits become less
constraining.  This is shown in Fig.~\ref{fig:NPrhoeta}, which shows the
determination of the unitarity triangle from tree-dominated decays only, which
are unlikely to be affected by new physics.  The plot on the left shows the
current fit results, while the constraints in the plot on the right is expected
to be achievable with 50\,ab$^{-1}$ Belle~II and 50\,fb$^{-1}$ LHCb
data~\cite{Charles:2013aka}.  The allowed region in the left plot is indeed
significantly larger than in Fig.~\ref{fig:SMCKMfit}.

\begin{figure}[tb]
\centerline{
\includegraphics[height=6cm,clip,bb=15 15 490 449]{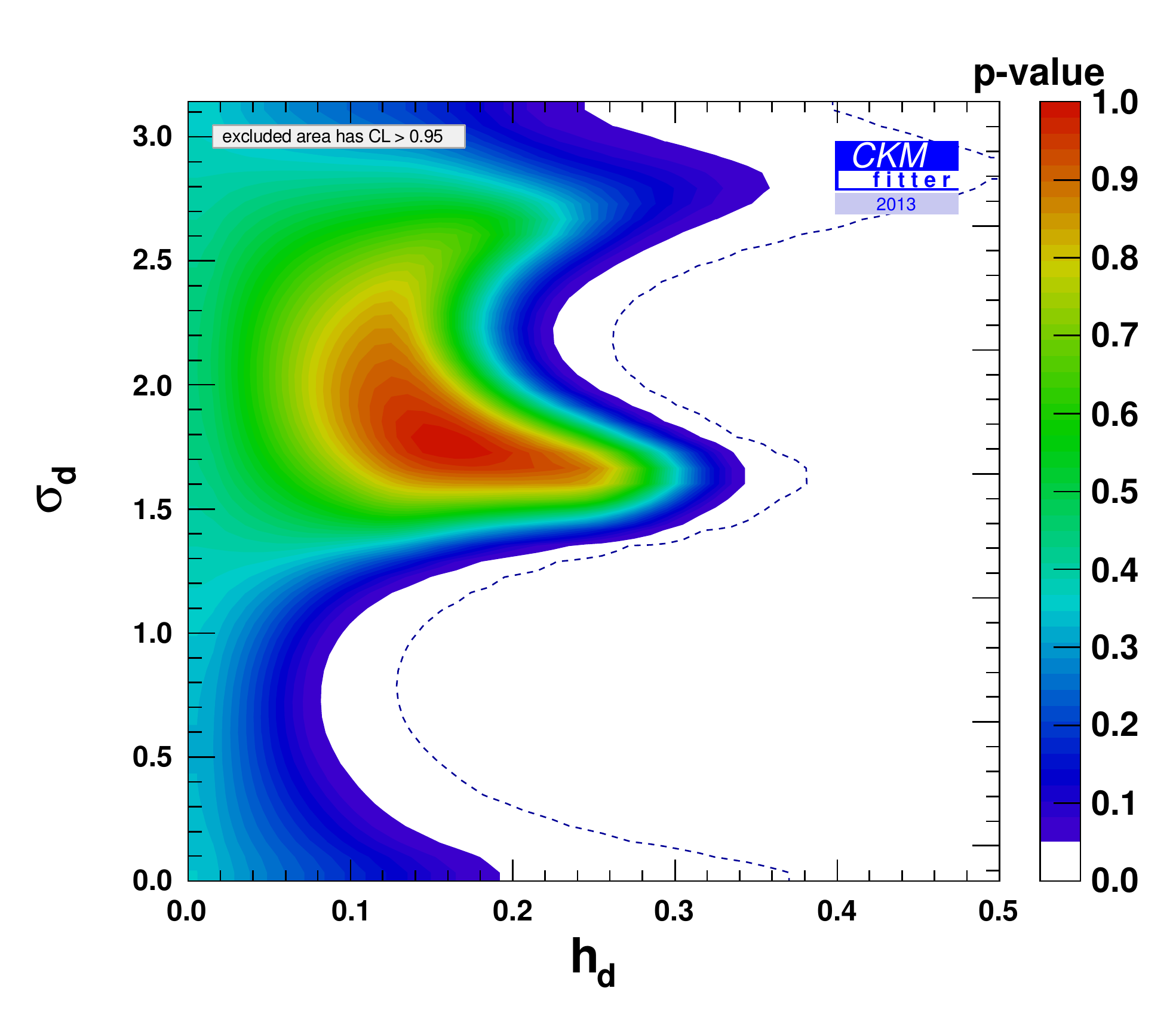}
\hfil\hfil
\includegraphics[height=6cm,clip,bb=15 15 550 449]{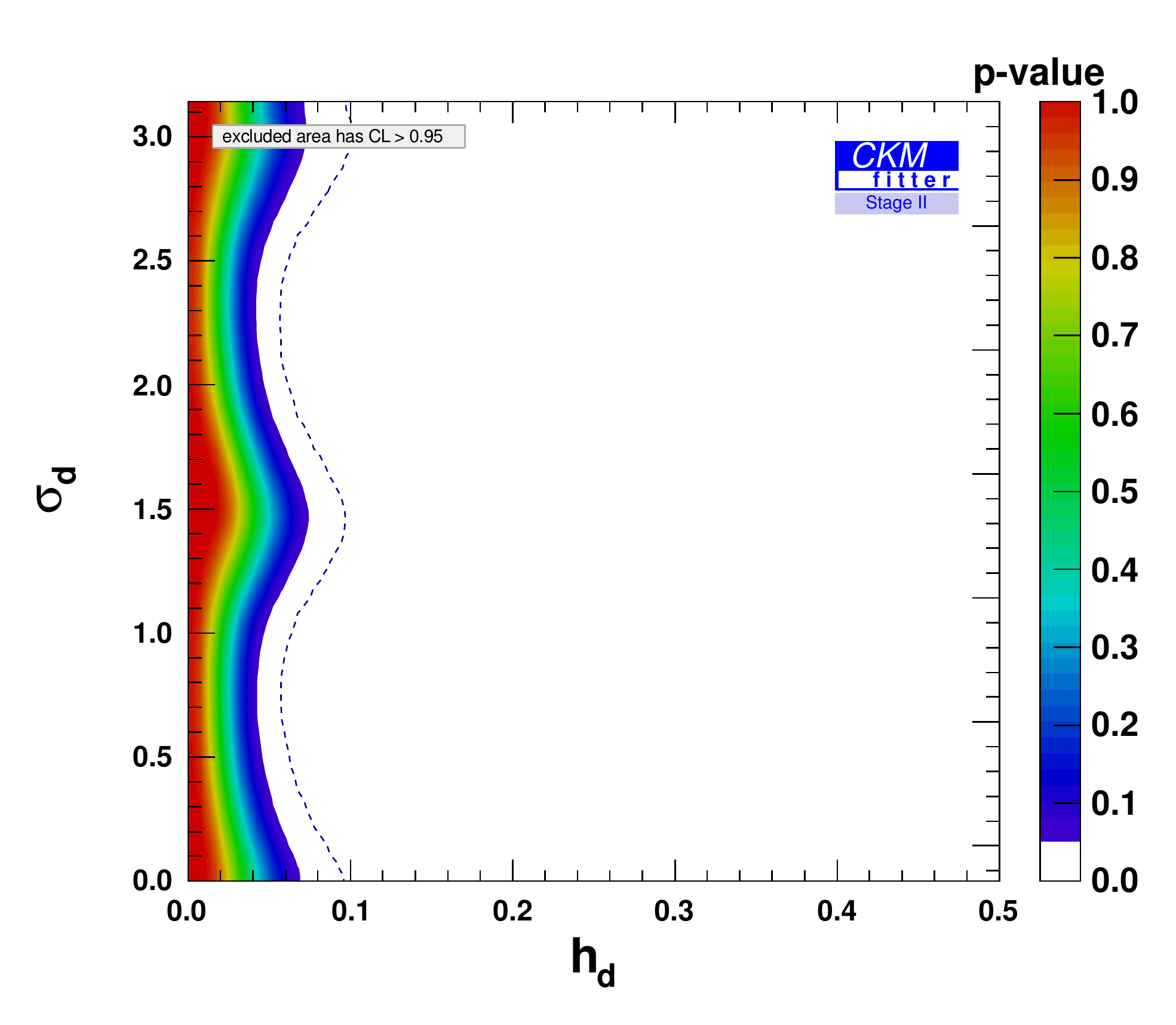}}
\caption{Constraints on the $h_d - \sigma_d$ parameters (left) and those
estimated to be achievable using 50\,ab$^{-1}$ Belle~II and 50\,fb$^{-1}$ LHCb
data (right)~\cite{Charles:2013aka}.  Colored regions show $2\sigma$ limits with
the colors indicating CL as shown, while the dashed curves show $3\sigma$
limits.}
\label{fig:NPBdmix}
\end{figure}

\begin{figure}[tb]
\centerline{
\includegraphics[height=6cm,clip,bb=15 15 490 449]{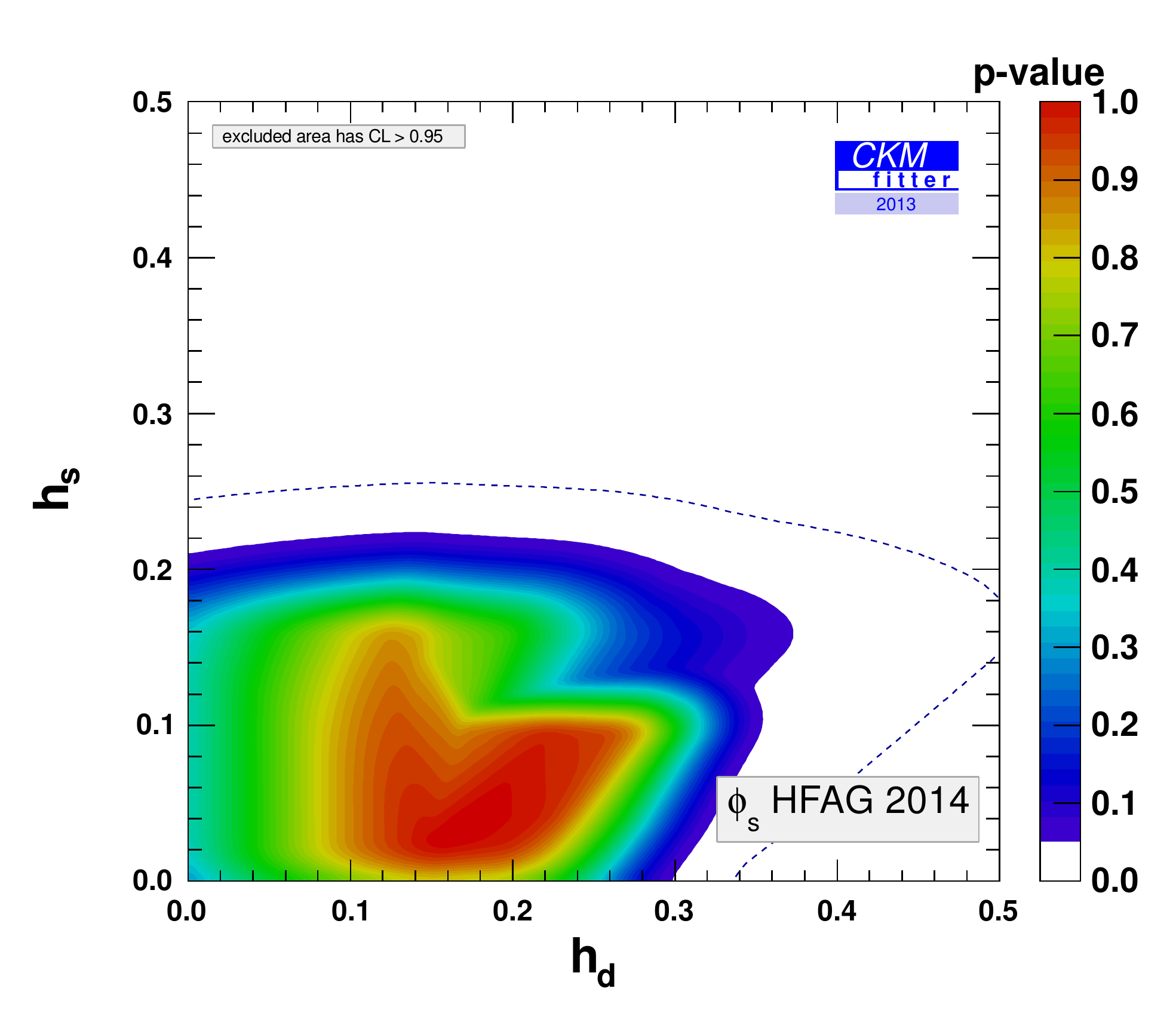}
\hfil\hfil
\includegraphics[height=6cm,clip,bb=15 15 550 449]{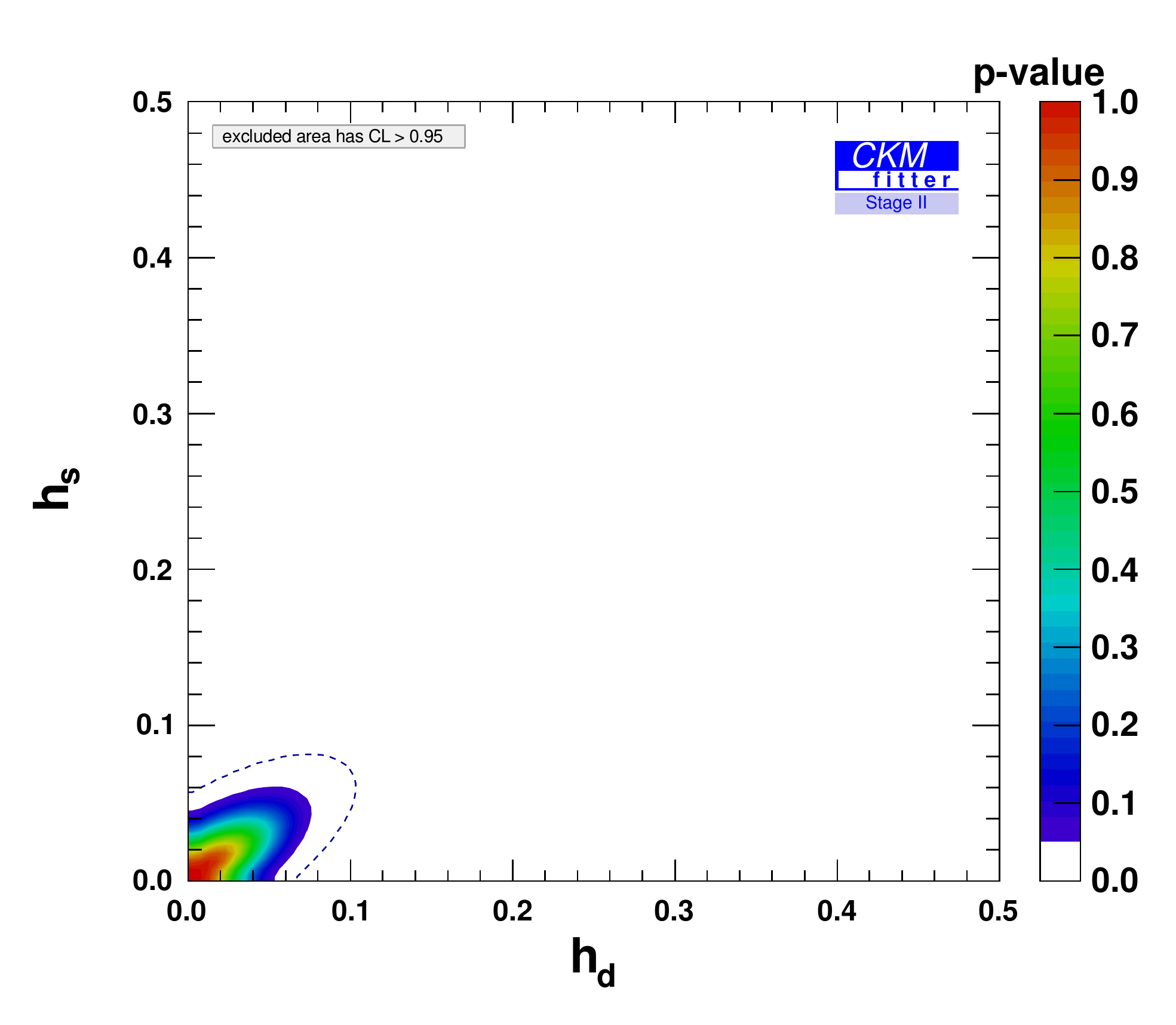}}
\caption{Constraints on the $h_d - h_s$ parameters now (left plot) and those
estimated with 50\,ab$^{-1}$ Belle~II and 50\,fb$^{-1}$ LHCb data (right
plot)~\cite{Charles:2013aka}.  The notation is the same as in
Fig.~\protect\ref{fig:NPBdmix}.}
\label{fig:NPBdBsmix}
\end{figure}

It has been known for decades that the mixing of neutral mesons is particularly
sensitive to new physics, and probe some of the highest scales.  In a large
class of models, NP has a negligible impact on tree-level SM transitions (e.g.,
the measurements of $\gamma$, $|V_{ub}|$, and $|V_{cb}|$), and the $3\times 3$
CKM matrix remains unitary.  As a simple example, consider possible NP
contributions to $B$ and $B_s$ meson mixing, which can be parametrized as
\beq\label{M12param}
M_{12} = M_{12}^{\rm SM} (1 + h_q\, e^{2i\sigma_q})\,, \qquad q=d,s\,.
\eeq
The constraints on $h_d$ and $\sigma_d$ in the $B_d$ mixing are shown in
Fig.~\ref{fig:NPBdmix}, and the constraint in the $h_d - h_s$ plane is shown in 
Fig.~\ref{fig:NPBdBsmix}.  Both plots show the current constraints (left) and
those expected to be achievable with 50\,ab$^{-1}$ Belle~II and 50\,fb$^{-1}$
LHCb data (right)~\cite{Charles:2013aka}.  Figure~\ref{fig:NPBdmix} shows that
in the future the bounds on the ``MFV-like regions", where NP flavor is aligned
with the SM ($2\theta_d \simeq 0$ mod $\pi$), will be comparable to  generic
values of the NP phase, unlike in the past.  Figure~\ref{fig:NPBdBsmix} shows
that the bounds on NP in $B_s$ mixing, which were significantly weaker than
those in the $B_d$ sector until recent LHCb measurements, are now comparable,
and will comparably improve in the future.

As an example, if NP modifies the SM operator describing $B_q$ mixing
by adding to it a term
\beq\label{SMmix}
\frac{C_{q}^2}{\Lambda^2}\, (\bar b_{L}\gamma^{\mu}q_{L})^2\,,
\eeq
then one finds
\beq
h_q \simeq \frac{|C_{q}|^2}{|V_{tb}^*\, V_{tq}|^2} 
  \left(\frac{4.5\, \TeV}{\Lambda}\right)^2\,.
\eeq
We can then translate the plotted bounds to the scale of new physics probed. 
The summary of expected sensitivities are shown in Table~\ref{scaletable}. The
sensitivities even with SM-like loop- and CKM-suppressed coefficients are
comparable to the scales probed by the LHC in the next decade.

\begin{table}[t]\tabcolsep 14pt
\renewcommand{\arraystretch}{1.2}
\centerline{
\begin{tabular}{c|c|c|c}
\hline\hline
\multirow{2}{*}{Couplings}  &  NP loop  &  
  \multicolumn{2}{c}{Scales (TeV) probed by}\\
&  order  &  $B_d$ mixing  &  $B_s$ mixing   \\
\hline
$|C_{q}| = |V_{tb}V_{tq}^*|$ & tree level &  17  &  19 \\
\cdashline{2-4}
(CKM-like)  &  one loop  &  1.4  &  1.5  \\
\hline
$|C_{q}| = 1$  &  tree level  &  $2\times 10^3$  &  $5\times 10^2$  \\
\cdashline{2-4}
(no hierarchy)  &  one loop &  $2\times 10^2$  &  40 \\
\hline\hline
\end{tabular}}
\caption{The scale of the $B_{d,s}$ mixing operators in
Eq.~(\protect\ref{SMmix}) probed, with 50\,ab$^{-1}$ Belle~II and 50\,fb$^{-1}$
LHCb data~\cite{Charles:2013aka}.  The differences due to CKM-like hierarchy of
couplings and/or loop suppression is shown.}
\label{scaletable}
\end{table}

In $K^0$\,--\,$\K0bar$ mixing the simplest analog of Eq.~(\ref{M12param}) is to
parameterize NP via an additive term to the so-called $tt$ contribution in the
SM, $M^{K,tt}_{12} = M^{K,tt}_{12} (1 + h_K\, e^{2i\sigma_K})$.  The reason is
the short distance nature of NP and the fact that in many NP models the largest
contribution to $M^K_{12}$ arise via effects involving the third generation. 
Substantial progress would require lattice QCD to constrain the long distance
contribution to $M_{12}^K$ at the percent level~\cite{Charles:2013aka}.

There are also strong constraints on NP from $D^0$\,--\,$\D0bar$ mixing.  Since
the observed mixing parameters are probably dominated by long distance
physics~\cite{Falk:2001hx}, it is hard to improve the bound from simply
demanding the NP contribution to be below the measured values of the mixing
parameters.

\subsection{Sensitivity to vector-like fermions}

Another illustration of the expected progress with well quantifiable increases
in mass scale sensitivity,  in both quark and lepton flavor experiments, is to
consider extensions of the SM involving vector-like fermions, which can Yukawa
couple to the SM~\cite{Ishiwata:2015cga}.  These fermions can have masses $M$
much greater than the weak scale, since they have a mass term even in the
absence of electroweak symmetry breaking.  These models are a class of simple
extensions of the SM, which do not worsen the hierarchy puzzle.  There are 11
renormalizable models~\cite{Ishiwata:2015cga} which add to the SM vector-like
fermions in a single (complex) representation of the gauge group that can Yukawa
couple to the SM fermions through the Higgs field (4 to leptons, 7 to quarks).

The precise definitions of the $\lambda_i$ Yukawa couplings depend on the
models, as do the forms of the Lagrangians.  For example, what was labeled
Model~V in Ref.~\cite{Ishiwata:2015cga} contains vector-like fermions, $D$, with
the same quantum numbers as the SM right-handed down-type quarks, which Yukawa
couple to the SM left-handed quark doublets $Q_L^i$ as
\beq
{\cal  L }_{\rm NP}^{\rm (V)} = {\bar D} (i\slashed{D} - M) D
  - (\lambda_i {\bar D_R} H^{\dagger} Q_L^i + {\rm h.c.} ) \,,
\eeq
These new interactions generate $Z$ couplings, e.g., in this Model~V to the
quarks,
\beq\label{bsmZ}
{\cal L}_Z^{\rm (V)} =
  - \sum_{i,j} \left( {\lambda_i^* \lambda_j\, m_Z^2 \over g_Z\, M^2}\right) 
  {\bar d}^{i}_L \gamma^{\mu} d^{j}_L\, Z_\mu \,,
\eeq
which contribute to, and are constrained by, flavor-changing neutral currents.

\begin{figure}[tb]
\centerline{\includegraphics[width=.3\textwidth]{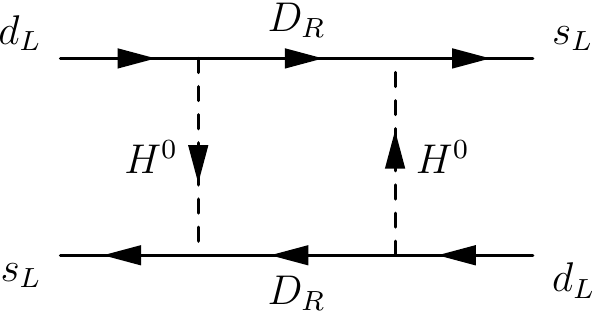} \hfil
\includegraphics[width=.3\textwidth]{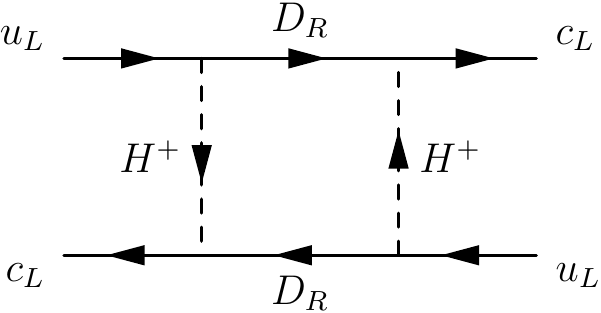}}
\caption{One-loop vector-like fermion contributions to $K$ and $D$ mixing in 
Model~V~\cite{Ishiwata:2015cga}.}
\label{fig:mixing}
\end{figure}

These models also generate dimension-6 four-fermion operators, which contribute
to neutral meson mixing.  At tree level, the $Z$ contribution in
Eq.~(\ref{bsmZ}) yields coefficients of the form $(\lambda_i \lambda_j^{*})^2
v^2/M^4$.  At one loop, coefficients of order $( \lambda_i \lambda_j^{*})^2 /(4
\pi M)^2$ are generated, which are neither CKM nor quark-mass suppressed,
seemingly not considered in the literature.  For large $M$, these one-loop
contributions are more important than tree-level $Z$ exchange.  They are
independent of the Higgs vacuum expectation value, $v$, and arise from short
distances $\sim 1/M$.  They can be calculated in the symmetric phase from the
box diagrams in Fig.~\ref{fig:mixing} with virtual scalars and the heavy
vector-like fermions.  The resulting effective Lagrangian in Model~V
is~\cite{Ishiwata:2015cga},
\beq
{\cal L}_{\rm mix}^{\rm (V)} = - {(\lambda_i^* \lambda_j )^2\over 128 \pi^2 M^2}
  \bigg[ \sum_{klmn} \big( {\bar u}_L^k V_{ki}\, \gamma_{\mu} V^{\dagger}_{jl}\,
  u_L^l \big) 
\big( {\bar u}_L^m V_{mi}\,\gamma_{\mu} V^{\dagger}_{jn}u_L^n \big)
  + \big( {\bar d}_L^i \gamma_{\mu}d_L^j \big)
  \big( {\bar d}_L^i \gamma^{\mu}d_L^j \big) \bigg] + {\rm h.c.}
\eeq

Table~\ref{tab:bounds} summarizes the bounds on 5 of the 11 models for
illustration.  The upper rows for each model show the current bounds, and the
lower rows show the expected sensitivities in the next generation of experiments
(in the next decade or so).  For the vector-like fermions that couple to SM
quarks, the bounds are  shown separately from $\Delta F=1$ and $\Delta F=2$
processes.  For the $\Delta F=2$ bounds on the 1--2 generation couplings, the
bounds are shown separately on the real and imaginary parts, since $\epsilon_K$
probes much higher scales than $\Delta m_K$ in these models.  (In the other
cases the differences are of order unity.)  We learn that the next generation
of experiments will improve the mass scale sensitivities in the leptonic
(hadronic) models by up to a factor of $\sim 7$ ($\sim 4$).

\begin{table}[tb] \tabcolsep 6pt
\centerline{\begin{tabular}{c@{\extracolsep{2pt}}c|cc|cc|cc}
\hline\hline
\multirow{2}{*}{Model} &  \multicolumn{1}{c|}{Quantum}  &
  \multicolumn{6}{c}{Bounds on $M/\TeV$ and $\lambda_i \lambda_j$
  for each $ij$ pair}  \\
& \multicolumn{1}{c|}{numbers} & \multicolumn{2}{c|}{$ij=12$} 
  & \multicolumn{2}{c|}{$ij=13$} & \multicolumn{2}{c}{$ij=23$}  \\
\hline
II & $(1,3,-1)$ & \multicolumn{2}{c|}{220$^a$} & \multicolumn{2}{c|}{4.9$^b$}
 & \multicolumn{2}{c}{5.2$^c$}
\\
 &  & \multicolumn{2}{c|}{1400$^a$} & \multicolumn{2}{c|}{13$^b$}
  & \multicolumn{2}{c}{15$^c$}
\\
III & $(1,2,-1/2)$ & \multicolumn{2}{c|}{310$^a$} & \multicolumn{2}{c|}{7.0$^b$}
  & \multicolumn{2}{c}{7.4$^c$}
\\
 &  & \multicolumn{2}{c|}{2000$^a$} & \multicolumn{2}{c|}{19$^b$}
  & \multicolumn{2}{c}{21$^c$} 
\\ \cline{3-8}
& & $\Delta F=1$ & $\Delta F=2$ & $\Delta F=1$ & $\Delta F=2$
  & $\Delta F=1$ & $\Delta F=2$
\\ \cline{3-8}
V & $(3,1,-1/3)$ & 66$^d$ [100]$^e$ & \{42, 670\}$^f$
  & 30$^g$ & 25$^h$ & 21$^i$ & 6.4$^j$
\\
 &  & 280$^d$ & \{$100$, 1000\}$^f$  & 60$^l$ & 61$^h$ & 39$^k$ & 14$^j$
\\
VII & $(3,3,-1/3)$  & 47$^d$ [71]$^e$ & \{47, 750\}$^f$ 
  & 21$^g$ & 28$^h$ & 15$^i$ & 7.2$^j$
\\
 &  & 200$^d$ & \{$110$, 1100\}$^f$  & 42$^l$ & 68$^h$ & 28$^k$ & 16$^j$
\\
XI & $(3,2,-5/6)$  & 66$^d$ [100]$^e$ & \{42, 670\}$^f$
  & 30$^g$ & 25$^h$ & 18$^k$ & 6.4$^j$
\\
 &  & 280$^d$ & \{$100$, 1000\}$^f$  & 60$^l$ & 61$^h$ & 39$^k$ & 14$^j$
\\
\hline\hline
\end{tabular}}
\caption{Bounds in some of the vector-like fermion
models~\cite{Ishiwata:2015cga} on $M\, [\TeV] /\sqrt{|\lambda_i \lambda_j|}$ in
the leptonic models, and from the $\Delta F=1$ constraints on the hadronic
models.  The $\Delta F=2$ bounds show $M/\sqrt{|\lambda_i\lambda_j|^2}$, except
for $K$ meson mixing we show $\big\{ M/\sqrt{|{\rm
Re}(\lambda_i\lambda_j^*)^2|}, \ M/\sqrt{|{\rm Im}(\lambda_i\lambda_j^*)^2|}\,
\big\}$.  The strongest bounds arise, or are expected to arise, from: $a)$ $\mu$
to $e$ conversion; $b)$ $\tau\to e\pi$; $c)$ $\tau\to \mu\rho$; $d)$
$K\to\pi\nu\bar\nu$; $e)$ $K_L\to\mu^+\mu^-$ (this involves $|{\rm
Re}(\lambda_1\lambda_2^*)|$ and is in square brackets because prospects for
improvements are weak); $f)$ $K$ mixing; $g)$ $B\to\pi\mu^+\mu^-$; $h)$ $B_d$
mixing; $i)$ $B\to X_s\ell^+\ell^-$; $j)$ $B_s$ mixing;  $k)$
$B_s\to\mu^+\mu^-$, $l)$ $B_d\to\mu^+\mu^-$.}
\label{tab:bounds}
\end{table}

\subsection{Top, higgs, and new physics flavor}

These are vast topics which I could not cover in detail in the talk, nor is it
possible here.  

Top quarks in the SM decay almost exclusively to $bW$, with the second largest
branching fraction ${\cal B}(t\to sW) < 2\times 10^{-3}$.  Particularly clean
probes of the SM are FCNC top decays, for which the SM predictions are below the
$10^{-12}$ level.  The current bounds are roughly at the level ${\cal B}(t\to
qZ) \lsim 10^{-3}$, ${\cal B}(t\to qg) \lsim 10^{-4}$, and ${\cal B}(t\to qh)
\lsim 0.5\%$, with the precise limits depending on the ratio of $q=u,c$ produced
by new physics~\cite{AMtalk}.  The ultimate LHC sensitivities are expected to be
about a factor of $10^2$ better, hence any observation would be a clear sign of
NP.  There is obvious complementarity between FCNC searches in the top sector,
and low energy flavor physics bounds.  Since $t_L$ is in the same $SU(2)$
doublet as $b_L$, several operators have correlated effects in $t$ and $b$
decays.  For some operators, mainly those involving left-handed quark fields,
the low energy constraints exclude a detectable LHC signal, whereas other
operators are still allowed to have large enough coefficients to yield
detectable NP signals at the LHC (see, e.g., Ref.~\cite{Fox:2007in}).

The experimental richness of higgs physics, that several production mechanisms
and many decay channels can be probed, are to a large extent due to the
particular values of the Yukawa couplings. The quark and lepton couplings, and
$Y_t$ in particular, are important for higgs decays, as well as to determine the
production cross sections from $gg$ fusion, higgs-strahlung, $t\bar t$ and $WZ$
fusion.  The LHC has (almost) measured the $h\tau^+\tau^-$ coupling, and will
also determine $h\mu^+\mu^-$ and $hb\bar b$, if they are near their SM values. 
Should the LHC or another future collider detect deviations from the SM
branching ratios or observe flavor-non-diagonal higgs decays, that would of
course be incredibly significant (for a recent discussion, see, e.g.,
Ref.~\cite{Nir:2016zkd}).

Any new particle that couples to the quarks and/or leptons, potentially
introduces new flavor violating parameters.  For example, in low energy
supersymmetry, which is the favorite NP scenario of a large part of our
community, squark and slepton couplings may yield measurable effects in FCNC
processes and $CP$ violation, give rise to detectable charged lepton flavor
violation (CLFV), such as $\mu\to e \gamma$, etc.  Observable $CP$ violation is
then also possible in neutral currents and electric dipole moments, for which
the SM predictions are below the near future experimental sensitivities.  The
supersymmetric flavor problems, that TeV-scale SUSY models with generic
parameters are excluded by FCNC and $CP$ violation measurements, can be
alleviated in several scenarios: (i) universal squark masses, when $\Delta\tilde
m_{\tilde Q,\tilde D}^2 \ll \tilde m^2$ (e.g., gauge mediation); (ii) alignment,
when $(K^d_{L,R})_{12} \ll 1$ (e.g., horizontal symmetry); (iii) heavy squarks,
when $\tilde m \gg 1\,\TeV$ (e.g., split SUSY).  All viable models incorporate
some of these ingredients.  Conversely, if SUSY is discovered, mapping out its
flavor structure may help answer questions about even higher scales, e.g., the
mechanism of SUSY breaking, how it is communicated to the MSSM, etc.

An important implication of flavor constraints for SUSY searches is that the LHC
bounds are sensitive to the level of (non-)degeneracy assumed.  Most SUSY
searches assume that the first two generation squarks, $\tilde u_{L,R}$, $\tilde
d_{L,R}$, $\tilde s_{L,R}$, $\tilde c_{L,R}$, are all degenerate, which
increases signal cross sections.  Relaxing this assumption consistent with
flavor bounds, results in substantially weaker squark mass limits from the LHC
Run~1, around the 500\,GeV scale~\cite{Mahbubani:2012qq}.  Thus, there is a
tight interplay between the flavor physics and LHC high-$p_T$ searches for new
physics.  If there is new physics at the TeV scale, its flavor structure must be
highly non-generic to satisfy current bounds, and measuring small deviations
from the SM in the flavor sector would give a lot of information complementary
to ATLAS \& CMS.  The higher the scale of new physics, the less severe the
flavor constraints are.  If NP is beyond the reach of the LHC, flavor physics
experiments may still observe robust deviations from the SM, which would point
to an upper bound on the next scale to probe.

\section{Final comments and ultimate sensitivity}
\label{sec:concl}

The main points I tried to convey through some examples were:

\begin{itemize}\vspace*{-6pt}\itemsep -2pt

\item $CP$ violation and FCNCs are sensitive probes of short-distance physics
in the SM and for~NP;

\item Flavor physics probes energy scales $\gg\!1\,\TeV$, the sensitivity
limited by statistics, not theory;

\item For most FCNC processes NP\,/\,SM $\gsim 20\%$ is still allowed, so there
is plenty of room for NP;

\item Of the several tensions between data and SM predictions, some may soon
become definitive;

\item Precision tests of SM will improve by $10^1$\,--\,$10^4$ in many channels
(including CLFV);

\item There are many interesting theory problems, relevant for improving
experimental sensitivity;

\item Future data will teach us more about physics at shorter distances,
whether NP is seen or not, and could point to the next energy scale to explore.

\end{itemize}\vspace*{-6pt}

With several new experiments starting (NA62, KOTO, Belle~II, mu2e, COMET, etc.)
and the upcoming upgrade of LHCb, the flood of new data will be fun and exciting
(see Refs.~\cite{YStalk, Bediaga:2012py, Belle2predictions} for reviews of
planned flavor experiments and their sensitivities).  It will allow new type of
measurements, and more elaborate theoretical methods to be used/tested.  The
upcoming experiments also challenge theory, to improve predictions and to allow
more measurements to probe short distance physics with robust discovery
potential.  Except for the few cleanest cases, improvements on both sides are
needed to fully exploit the future data sets.  I am optimistic, as order of
magnitude increases in data always triggered new theory developments, too.

It is also interesting to try to estimate the largest flavor physics data sets
which would be useful to increase sensitivity to new physics, without being
limited by theory uncertainties.\footnote{In measurements without SM
backgrounds, such as setting bounds on $\mu\to e$ conversion or $\tau\to 3\mu$
decay, the mass-scale sensitivity (to a dimension-6 NP operator) scales like
$\Lambda \propto (\mbox{bound})^{-1/4}$.  In measurements constraining SM--NP
interference, $\Lambda \propto (\mbox{uncertainty})^{-1/2}$, and at some point
precise knowledge of the SM contribution becomes critical.}  For charged lepton
flavor violation, the SM predictions (from penguin and box diagrams with
neutrinos) are (tens of) orders of magnitudes below any foreseeable experimental
sensitivity, so if technology allows significant improvements, I think the
justification is obvious (as it is for electric dipole moment searches).  In
quark flavor physics the situation is more complex.  Amusingly, even in 2030,
there will be theoretically clean $B$ decay modes in which (experimental
bound)\big/SM $\gtrsim 10^3$, e.g., $B\to \tau^+\tau^-$, $B\to e^+e^-$, and
probably some more.  However, based on what is known today, some observables
will become limited by theory (hadronic) uncertainties.  Identifying how far NP
sensitivity can be extended is interesting, at least in principle, so below is a
list for which 50/ab Belle~II and 50/fb LHCb data will not even come within an
order of magnitude of the ultimately achievable sensitivities.  Of course, on
the relevant time scale lots of progress will take place (for estimates of
future lattice QCD uncertainties, see, Ref.~\cite{Butler:2013kdw}) and new
breakthroughs are also possible.  

\begin{itemize}\vspace*{-6pt}\itemsep -2pt

\item Probably the theoretically cleanest observable in the quark sector is the
determination of the CKM phase $\gamma$ from tree-level $B$ decays.  Irreducible
theory uncertainty only arises from higher order weak
interaction~\cite{gammaTH}.  So the main challenges are on the experimental
side.  

\item The theory uncertainty for the semileptonic $CP$ asymmetries, $a_{\rm
SL}^{d,s}$, discussed is Sec.~\ref{sec:status} and in Fig~\ref{fig:cpvmix}, are
also much below~\cite{Laplace:2002ik, Lenz:2011ti} the expected 50/ab Belle~II
and 50/fb LHCb sensitivities.

\item Another set of key observables are $B_{s,d}\to\mu\mu$ and $B\to\ell\nu$,
where the nonperturbative theory inputs are only the decay constants, which will
soon be known with $<1\%$ uncertainties.  In contrast, the expectation for the
accuracy of $B_d\to\mu\mu$ with the full LHC data is ${\cal O}(20\%)$.

\item It is often stated that the determination of $|V_{ub}|$ is theory
limited.  This entirely depends on the measurements available.  In principle,
the theoretically cleanest $|V_{ub}|$ determination I know, which only uses
isospin, would be from ${\cal B}(B_u\to \ell\bar\nu) / {\cal B}(B_d\to
\mu^+\mu^-)$~\cite{BGckm06}.

\item I think that the SM prediction for $CP$ violation in $D^0$\,--\,$\D0bar$
mixing is below the expected sensitivities on LHCb and Belle~II.  To establish
this robustly, however, more theory work is needed (especially given the recent
history of hints of $CP$ violation in $D$ decay).

\item For $K^+\to\pi^+\nu\bar\nu$ and especially for $K_L\to\pi^0\nu\bar\nu$,
the current plans for NA62 and KOTO will stop short of reaching the ultimate
sensitivity to NP achievable in these decays.

\end{itemize}\vspace*{-6pt}

Thus, I guess(timate) that $\sim 100$ times the currently envisioned 50/ab
Belle~II and 50/fb LHCb data sets would definitely allow for the sensitivity to
short distance physics to improve.  Whether any of these ultimate sensitivities
can be achieved at a tera-$Z$ machine, an $e^+e^-$ collider running on the
$\Upsilon(4S)$, or utilizing more of the LHC's or/and a future hadron collider's
full luminosity, is something I hope we shall soon have even more compelling
reasons to seriously explore.

\medskip
\paragraph{Acknowledgments}
I thank Marat Freytsis, Tim Gershon, Koji Ishiwata, Michele Papucci, Dean
Robinson, Josh Ruderman, Filippo Sala, Karim Trabelsi, Phill Urquijo, and Mark
Wise, for recent collaborations and/or discussions that shaped the views
expressed in this talk. This work was supported in part by the Director, Office
of Science, Office of High Energy Physics of the U.S.\ Department of Energy
under contract DE-AC02-05CH11231.


\begin{thebibliography}{99}
%%\addcontentsline{toc}{section}{References}

\itemsep 1pt plus 1pt minus 1pt %% default is 3+2-1
%\bigskip\the\itemsep

\bibitem{Abdesselam:2016cgx}
  A.~Abdesselam {\it et al.} [Belle Collaboration],
  %``Measurement of the branching ratio of $\bar{B}^0 \rightarrow D^{*+} \tau^- \bar{\nu}_{\tau}$ relative to $\bar{B}^0 \rightarrow D^{*+} \ell^- \bar{\nu}_{\ell}$ decays with a semileptonic tagging method,''
  \href{http://arXiv.org/abs/1603.06711}{arXiv:1603.06711}.
  %%CITATION = ARXIV:1603.06711;%%

\bibitem{Abdesselam:2016llu} 
  A.~Abdesselam {\it et al.} [Belle Collaboration],
  %``Angular analysis of $B^0 \to K^\ast(892)^0 \ell^+ \ell^-$,''
  \href{http://arXiv.org/abs/1604.04042}{arXiv:1604.04042}.
  %%CITATION = ARXIV:1604.04042;%%

\bibitem{Aaboud:2016ire}
  M.~Aaboud {\it et al.} [ATLAS Collaboration],
  %``Study of the rare decays of $B^0_s$ and $B^0$ into muon pairs from data collected during the LHC Run 1 with the ATLAS detector,''
  \href{http://arXiv.org/abs/1604.04263}{arXiv:1604.04263}.

\bibitem{Aaij:2016yze} 
  R.~Aaij {\it et al.} [LHCb Collaboration],
  %``Measurement of the $CP$ asymmetry in $B_s^0-\overline{B}{}_s^0$ mixing,''
  \href{http://arXiv.org/abs/1605.09768}{arXiv:1605.09768}.

\bibitem{Hocker:2001xe}
  A.~H\"ocker, H.~Lacker, S.~Laplace and F.~Le Diberder,
%  ``A New approach to a global fit of the CKM matrix,''
  Eur.\ Phys.\ J.\ C {\bf 21}, 225 (2001)
  [\href{http://arXiv.org/abs/hep-ph/0104062}{hep-ph/0104062}];
  %%CITATION = HEP-PH/0104062;%%
%\bibitem{Charles:2004jd}
  J.~Charles {\it et al.},
%  ``$CP$ violation and the CKM matrix: Assessing the impact of the asymmetric
%  $B$ factories,''
  Eur.\ Phys.\ J.\ C {\bf 41} (2005) 1
  [\href{http://arXiv.org/abs/hep-ph/0406184}{hep-ph/0406184}];
  %%CITATION = HEP-PH 0406184;%%
and updates at \href{http://ckmfitter.in2p3.fr/}{http://ckmfitter.in2p3.fr/}.

\bibitem{Ligeti:2015kwa}
  Z.~Ligeti,
  %``TASI Lectures on Flavor Physics,''
  \href{http://arXiv.org/abs/1502.01372}{arXiv:1502.01372}.
  %%CITATION = ARXIV:1502.01372;%%

\bibitem{PDG}
See the reviwes on the ``Cabibbo-Kobayashi-Maskawa quark mixing matrix" and on
``$CP$ violation in the quark sector" in:
  K.~A.~Olive {\it et al.}  [Particle Data Group],
  %``Review of Particle Physics,''
  Chin.\ Phys.\ C {\bf 38}, 090001 (2014).
  %%CITATION = CHPHD,C38,090001;%%

\bibitem{KMtalk}
K.~Miyabayashi, talk at this conference,
\href{http://indico.cern.ch/event/325831/contributions/757780/}
{http://indico.cern.ch/event/325831/contributions/757780/}.

\bibitem{Lees:2012xj}
  J.~P.~Lees {\it et al.}  [BaBar Collaboration],
  %``Evidence for an excess of $\bar{B} \to D^{(*)}\tau^-\bar{\nu}_\tau$ decays,''
  Phys.\ Rev.\ Lett.\  {\bf 109}, 101802 (2012)
  [\href{http://arXiv.org/abs/1205.5442}{arXiv:1205.5442}].
  %%CITATION = ARXIV:1205.5442;%%

\bibitem{Lees:2013udz}
  J.~P.~Lees {\it et al.}  [BABAR Collaboration],
  %``Measurement of an Excess of B -> D(*) Tau Nu Decays and Implications for Charged Higgs Bosons,''
  Phys.\ Rev.\ D {\bf 88}, 072012 (2013)
  [\href{http://arXiv.org/abs/1303.0571}{arXiv:1303.0571}].
  %%CITATION = ARXIV:1303.0571;%%

\bibitem{Huschle:2015rga} 
  M.~Huschle {\it et al.} [Belle Collaboration],
  %``Measurement of the branching ratio of $\bar{B} \to D^{(\ast)} \tau^- \bar{\nu}_\tau$ relative to $\bar{B} \to D^{(\ast)} \ell^- \bar{\nu}_\ell$ decays with hadronic tagging at Belle,''
  Phys.\ Rev.\ D {\bf 92}, no. 7, 072014 (2015)
%  doi:10.1103/PhysRevD.92.072014
  [\href{http://arXiv.org/abs/1507.03233}{arXiv:1507.03233}].
  %%CITATION = doi:10.1103/PhysRevD.92.072014;%%

\bibitem{Aaij:2015yra} 
  R.~Aaij {\it et al.} [LHCb Collaboration],
  %``Measurement of the ratio of branching fractions $\mathcal{B}(\bar{B}^0 \to D^{*+}\tau^{-}\bar{\nu}_{\tau})/\mathcal{B}(\bar{B}^0 \to D^{*+}\mu^{-}\bar{\nu}_{\mu})$,''
  Phys.\ Rev.\ Lett.\  {\bf 115}, no. 11, 111803 (2015)
  Addendum: [Phys.\ Rev.\ Lett.\  {\bf 115}, no. 15, 159901 (2015)]
%  doi:10.1103/PhysRevLett.115.159901, 10.1103/PhysRevLett.115.111803
  [\href{http://arXiv.org/abs/1506.08614}{arXiv:1506.08614}].
  %%CITATION = doi:10.1103/PhysRevLett.115.159901, 10.1103/PhysRevLett.115.111803;%%

\bibitem{HFAG}
  Heavy Flavor Averaging Group, Y.~Amhis
%  S.~Banerjee, E.~Ben-Haim, S.~Blyth, A.~Bozek, C.~Bozzi and A.~Carbone
  {\it et al.},
%  ``Averages of $b$-hadron, $c$-hadron, and $\tau$-lepton properties as of summer 2014,''
  \href{http://arXiv.org/abs/1412.7515}{arXiv:1412.7515};
  %%CITATION = ARXIV:1412.7515;%%
  and updates at
  \href{http://www.slac.stanford.edu/xorg/hfag/}%
  {http://www.slac.stanford.edu/xorg/hfag/}.

\bibitem{Lattice:2015rga} 
  J.~A.~Bailey {\it et al.} [MILC Collaboration],
  %``B\u2192D\u2113\u03bd form factors at nonzero recoil and |V$_{cb}$| from 2+1-flavor lattice QCD,''
  Phys.\ Rev.\ D {\bf 92}, no. 3, 034506 (2015)
%  doi:10.1103/PhysRevD.92.034506
  [\href{http://arXiv.org/abs/1503.07237}{arXiv:1503.07237}].
  %%CITATION = doi:10.1103/PhysRevD.92.034506;%%

\bibitem{Na:2015kha} 
  H.~Na {\it et al.} [HPQCD Collaboration],
  %``B\u2192Dl\u03bd form factors at nonzero recoil and extraction of |V$_{cb}$|,''
  Phys.\ Rev.\ D {\bf 92}, no. 5, 054510 (2015)
%  doi:10.1103/PhysRevD.92.054510
  [\href{http://arXiv.org/abs/1505.03925}{arXiv:1505.03925}].
  %%CITATION = doi:10.1103/PhysRevD.92.054510;%%

\bibitem{Fajfer:2012vx}
  S.~Fajfer, J.~F.~Kamenik and I.~Nisandzic,
  %``On the $B \to D* \tau \bar{nu)_\tau$ Sensitivity to New Physics,''
  Phys.\ Rev.\ D {\bf 85}, 094025 (2012)
  [\href{http://arXiv.org/abs/1203.2654}{arXiv:1203.2654}].
  %%CITATION = ARXIV:1203.2654;%%

\bibitem{Belle2predictions}
  B.~Golob, K.~Trabelsi, P.~Urquijo, BELLE2-NOTE-0021,
\href{https://belle2.cc.kek.jp/~twiki/pub/B2TiP/WebHome/belle2-note-0021.pdf}%
  {https://belle2.cc.kek.jp/~twiki/pub/B2TiP/WebHome/belle2-note-0021.pdf}.

\bibitem{Freytsis:2015qca}
  M.~Freytsis, Z.~Ligeti and J.~T.~Ruderman,
  %``Flavor models for $\bar{B} \to D^{(*)} \tau \bar{\nu}$,''
  Phys.\ Rev.\ D {\bf 92}, no. 5, 054018 (2015)
  %doi:10.1103/PhysRevD.92.054018
  [\href{http://arXiv.org/abs/1506.08896}{arXiv:1506.08896}].
  %%CITATION = doi:10.1103/PhysRevD.92.054018;%%

\bibitem{Varzielas:2015iva}
  I.~de Medeiros Varzielas and G.~Hiller,
  %``Clues for flavor from rare lepton and quark decays,''
  JHEP {\bf 1506}, 072 (2015)
  [\href{http://arXiv.org/abs/1503.01084}{arXiv:1503.01084}].
  %%CITATION = ARXIV:1503.01084;%%

\bibitem{future}
M.\ Freytsis, Z.\ Ligeti, J.\ Ruderman, to appear.

\bibitem{Ligeti:2014kia}
  Z.~Ligeti and F.~J.~Tackmann,
  %``Precise predictions for $B \to X_c \tau \bar \nu$ decay distributions,''
  Phys.\ Rev.\ D {\bf 90}, no. 3, 034021 (2014)
  [\href{http://arXiv.org/abs/1406.7013}{arXiv:1406.7013}];\\
  %%CITATION = ARXIV:1406.7013;%%
%\bibitem{Falk:1994gw} 
  A.~F.~Falk, Z.~Ligeti, M.~Neubert and Y.~Nir,
  %``Heavy quark expansion for the inclusive decay anti-B ---> tau anti-neutrino X,''
  Phys.\ Lett.\ B {\bf 326}, 145 (1994)
%  doi:10.1016/0370-2693(94)91206-8
  [\href{http://arXiv.org/abs/hep-ph/9401226}{hep-ph/9401226}].
  %%CITATION = doi:10.1016/0370-2693(94)91206-8;%%

\bibitem{Aubert:2008yv}
  B.~Aubert {\it et al.}  [BaBar Collaboration],
  %``Measurements of the Semileptonic Decays anti-B ---> D l anti-nu and anti-B ---> D* l anti-nu Using a Global Fit to D X l anti-nu Final States,''
  Phys.\ Rev.\ D {\bf 79}, 012002 (2009)
  [\href{http://arXiv.org/abs/0809.0828}{arXiv:0809.0828}].
  %%CITATION = ARXIV:0809.0828;%%

\bibitem{Dungel:2010uk}
  W.~Dungel {\it et al.}  [Belle Collaboration],
  %``Measurement of the form factors of the decay B0 -> D*- ell+ nu and determination of the CKM matrix element |Vcb|,''
  Phys.\ Rev.\ D {\bf 82}, 112007 (2010)
  [\href{http://arXiv.org/abs/1010.5620}{arXiv:1010.5620}].
  %%CITATION = ARXIV:1010.5620;%%

\bibitem{Greljo:2015mma}
  A.~Greljo, G.~Isidori and D.~Marzocca,
  %``On the breaking of Lepton Flavor Universality in B decays,''
  JHEP {\bf 1507}, 142 (2015)
  %doi:10.1007/JHEP07(2015)142
  [\href{http://arXiv.org/abs/1506.01705}{arXiv:1506.01705}].
  %%CITATION = doi:10.1007/JHEP07(2015)142;%%

\bibitem{Fortes:2015jaa} 
  E.~C.~F.~S.~Fortes and S.~Nussinov,
  %``Semitauonic B decay anomaly,''
  Phys.\ Rev.\ D {\bf 93}, 014023 (2016)
%  doi:10.1103/PhysRevD.93.014023
  [\href{http://arXiv.org/abs/1508.04463}{arXiv:1508.04463}].

\bibitem{Bernlochner:2015mya}
  F.~U.~Bernlochner,
  %``$B \to \pi \tau \overline \nu_\tau$ decay in the context of type II 2HDM,''
  Phys.\ Rev.\ D {\bf 92}, no. 11, 115019 (2015)
%  doi:10.1103/PhysRevD.92.115019
  [\href{http://arXiv.org/abs/1509.06938}{arXiv:1509.06938}].
  %%CITATION = doi:10.1103/PhysRevD.92.115019;%%

\bibitem{Hamer:2015jsa}
  P.~Hamer {\it et al.} [Belle Collaboration],
  %``Search for $B^0 \to \pi^- \tau^+ \nu_\tau$ with hadronic tagging at Belle,''
  Phys.\ Rev.\ D {\bf 93}, no. 3, 032007 (2016)
%  doi:10.1103/PhysRevD.93.032007
  [\href{http://arXiv.org/abs/1509.06521}{arXiv:1509.06521}].
  %%CITATION = doi:10.1103/PhysRevD.93.032007;%%

\bibitem{Descotes-Genon:2013wba} 
  S.~Descotes-Genon, J.~Matias and J.~Virto,
  %``Understanding the $B\to K^*\mu^+\mu^-$ Anomaly,''
  Phys.\ Rev.\ D {\bf 88}, 074002 (2013)
%  doi:10.1103/PhysRevD.88.074002
  [\href{http://arXiv.org/abs/1307.5683}{arXiv:1307.5683}].
  %%CITATION = doi:10.1103/PhysRevD.88.074002;%%

\bibitem{Altmannshofer:2014rta} 
  W.~Altmannshofer and D.~M.~Straub,
  %``New physics in $b\rightarrow s$ transitions after LHC run 1,''
  Eur.\ Phys.\ J.\ C {\bf 75}, no. 8, 382 (2015)
%  doi:10.1140/epjc/s10052-015-3602-7
  [\href{http://arXiv.org/abs/1411.3161}{arXiv:1411.3161}].
  %%CITATION = doi:10.1140/epjc/s10052-015-3602-7;%%

\bibitem{Aaij:2015oid}
  R.~Aaij {\it et al.} [LHCb Collaboration],
  %``Angular analysis of the $B^{0} \to K^{*0} \mu^{+} \mu^{-}$ decay using 3 fb$^{-1}$ of integrated luminosity,''
  JHEP {\bf 1602}, 104 (2016)
%  doi:10.1007/JHEP02(2016)104
  [\href{http://arXiv.org/abs/1512.04442}{arXiv:1512.04442}].
  %%CITATION = doi:10.1007/JHEP02(2016)104;%%

\bibitem{GLtalk}
G.~Lanfranchi, talk at this conference,
\href{http://indico.cern.ch/event/325831/contributions/757779/}
{http://indico.cern.ch/event/325831/contributions/757779/}.

\bibitem{Descotes-Genon:2014uoa}
  S.~Descotes-Genon, L.~Hofer, J.~Matias and J.~Virto,
  %``On the impact of power corrections in the prediction of $B \to K^*\mu^+\mu^-$ observables,''
  JHEP {\bf 1412}, 125 (2014)
%  doi:10.1007/JHEP12(2014)125
  [\href{http://arXiv.org/abs/1407.8526}{arXiv:1407.8526}].
  %%CITATION = doi:10.1007/JHEP12(2014)125;%%

\bibitem{Bauer:2002aj} 
  C.~W.~Bauer, D.~Pirjol and I.~W.~Stewart,
  %``Factorization and endpoint singularities in heavy to light decays,''
  Phys.\ Rev.\ D {\bf 67}, 071502 (2003)
%  doi:10.1103/PhysRevD.67.071502
  [\href{http://arXiv.org/abs/hep-ph/0211069}{hep-ph/0211069}].
  %%CITATION = doi:10.1103/PhysRevD.67.071502;%%

\bibitem{Beneke:2003pa}
  M.~Beneke and T.~Feldmann,
  %``Factorization of heavy to light form-factors in soft collinear effective theory,''
  Nucl.\ Phys.\ B {\bf 685}, 249 (2004)
%  doi:10.1016/j.nuclphysb.2004.02.033
  [\href{http://arXiv.org/abs/hep-ph/0311335}{hep-ph/0311335}].
  %%CITATION = doi:10.1016/j.nuclphysb.2004.02.033;%%

\bibitem{Charles:1998dr}
  J.~Charles, A.~Le Yaouanc, L.~Oliver, O.~Pene and J.~C.~Raynal,
  %``Heavy to light form-factors in the heavy mass to large energy limit of QCD,''
  Phys.\ Rev.\ D {\bf 60}, 014001 (1999)
%  doi:10.1103/PhysRevD.60.014001
  [\href{http://arXiv.org/abs/hep-ph/9812358}{hep-ph/9812358}].
  %%CITATION = doi:10.1103/PhysRevD.60.014001;%%

\bibitem{Jager:2014rwa} 
  S.~Jäger and J.~Martin Camalich,
  %``Reassessing the discovery potential of the $B \to K^{*} \ell^+\ell^-$ decays in the large-recoil region: SM challenges and BSM opportunities,''
  Phys.\ Rev.\ D {\bf 93}, no. 1, 014028 (2016)
%  doi:10.1103/PhysRevD.93.014028
  [\href{http://arXiv.org/abs/1412.3183}{arXiv:1412.3183}].
  %%CITATION = doi:10.1103/PhysRevD.93.014028;%%

\bibitem{Ciuchini:2015qxb}
  M.~Ciuchini, M.~Fedele, E.~Franco, S.~Mishima, A.~Paul, L.~Silvestrini and M.~Valli,
  %``$B\to K^* \ell^+ \ell^-$ decays at large recoil in the Standard Model: a theoretical reappraisal,''
  \href{http://arXiv.org/abs/1512.07157}{arXiv:1512.07157}.
  %%CITATION = ARXIV:1512.07157;%%

\bibitem{Gudrun}
G.~Hiller, Plenary talk at EPS 2015,
\href{https://indico.cern.ch/event/356420/contributions/1764018/}
{https://indico.cern.ch/event/356420/contributions/1764018/}.

\bibitem{Bobeth:2012vn} 
  C.~Bobeth, G.~Hiller and D.~van Dyk,
  %``General analysis of $\bar{B} \to \bar{K}^{(*)}\ell^+ \ell^-$  decays at low recoil,''
  Phys.\ Rev.\ D {\bf 87}, no. 3, 034016 (2013)
%  [Phys.\ Rev.\ D {\bf 87}, 034016 (2013)]
%  doi:10.1103/PhysRevD.87.034016
  [\href{http://arXiv.org/abs/1212.2321}{arXiv:1212.2321}].
  %%CITATION = doi:10.1103/PhysRevD.87.034016;%%

\bibitem{Ligeti:2007sn}
  Z.~Ligeti and F.~J.~Tackmann,
  %``Precise predictions for B ---> X(s) l+ l- in the large q**2 region,''
  Phys.\ Lett.\ B {\bf 653}, 404 (2007)
  [\href{http://arXiv.org/abs/0707.1694}{arXiv:0707.1694}];
  %%CITATION = ARXIV:0707.1694;%%
%\bibitem{Lee:2006gs}
  K.~S.~M.~Lee, Z.~Ligeti, I.~W.~Stewart and F.~J.~Tackmann,
  %``Extracting short distance information from b ---> s l+ l- effectively,''
  Phys.\ Rev.\ D {\bf 75}, 034016 (2007)
  [\href{http://arXiv.org/abs/hep-ph/0612156}{hep-ph/0612156}].
  %%CITATION = HEP-PH/0612156;%%

\bibitem{Aaij:2015esa}
  R.~Aaij {\it et al.} [LHCb Collaboration],
  %``Angular analysis and differential branching fraction of the decay $B^0_s\to\phi\mu^+\mu^-$,''
  JHEP {\bf 1509}, 179 (2015)
%  doi:10.1007/JHEP09(2015)179
  [\href{http://arXiv.org/abs/1506.08777}{arXiv:1506.08777}].
  %%CITATION = doi:10.1007/JHEP09(2015)179;%%

\bibitem{Straub:2015ica} 
  A.~Bharucha, D.~M.~Straub and R.~Zwicky,
  %``$B\to V\ell^+\ell^-$ in the Standard Model from Light-Cone Sum Rules,''
  \href{http://arXiv.org/abs/1503.05534}{arXiv:1503.05534}.
  %%CITATION = ARXIV:1503.05534;%%

\bibitem{Horgan:2015vla}
  R.~R.~Horgan, Z.~Liu, S.~Meinel and M.~Wingate,
  %``Rare $B$ decays using lattice QCD form factors,''
  PoS LATTICE {\bf 2014}, 372 (2015)
  [\href{http://arXiv.org/abs/1501.00367}{arXiv:1501.00367}].
  %%CITATION = ARXIV:1501.00367;%%

\bibitem{CMS:2014xfa}
  V.~Khachatryan {\it et al.} [CMS and LHCb Collaborations],
  %``Observation of the rare $B^0_s\to\mu^+\mu^-$ decay from the combined analysis of CMS and LHCb data,''
  Nature {\bf 522}, 68 (2015)
%  doi:10.1038/nature14474
  [\href{http://arXiv.org/abs/1411.4413}{arXiv:1411.4413}].
  %%CITATION = doi:10.1038/nature14474;%%

\bibitem{Abazov:2011yk}
  V.~M.~Abazov {\it et al.} [D0 Collaboration],
  %``Measurement of the anomalous like-sign dimuon charge asymmetry with 9 fb${^-1}$ of $p\bar{p}$ collisions,''
  Phys.\ Rev.\ D {\bf 84}, 052007 (2011)
  [\href{http://arXiv.org/abs/1106.6308}{arXiv:1106.6308}].
  %%CITATION = ARXIV:1106.6308;%%

\bibitem{Ligeti:2010ia}
  Z.~Ligeti, M.~Papucci, G.~Perez and J.~Zupan,
  %``Implication s of the dimuon CP asymmetry in B_{d,s} decays,''
  Phys.\ Rev.\ Lett.\  {\bf 105}, 131601 (2010)
  [\href{http://arXiv.org/abs/1006.0432}{arXiv:1006.0432}].
  %%CITATION = ARXIV:1006.0432;%%

\bibitem{Bernlochner:2014ova}
  F.~U.~Bernlochner, Z.~Ligeti and S.~Turczyk,
  %``New ways to search for right-handed current in B\u2192\u03c1\u2113$\bar{\u03bd}$ decay,''
  Phys.\ Rev.\ D {\bf 90}, no. 9, 094003 (2014)
  [\href{http://arXiv.org/abs/1408.2516}{arXiv:1408.2516}].
  %%CITATION = ARXIV:1408.2516;%%

\bibitem{deVries:2015hva}
  K.~J.~de Vries {\it et al.},
  %``The pMSSM10 after LHC Run 1,''
  Eur.\ Phys.\ J.\ C {\bf 75}, no. 9, 422 (2015)
  [\href{http://arXiv.org/abs/1504.03260}{arXiv:1504.03260}].
  %%CITATION = ARXIV:1504.03260;%%

\bibitem{MPtalk}
M.~Pospelov, talk at this conference,
\href{http://indico.cern.ch/event/325831/contributions/757807/}
{http://indico.cern.ch/event/325831/contributions/757807/}.

\bibitem{Ligeti:2016qpi}
  Z.~Ligeti and F.~Sala,
  %``A new look at the theory uncertainty of epsilon_K,''
  \href{http://arXiv.org/abs/1602.08494}{arXiv:1602.08494}.
  %%CITATION = ARXIV:1602.08494;%%

\bibitem{Bai:2015nea}
  Z.~Bai {\it et al.} [RBC and UKQCD Collaborations],
  %``Standard Model Prediction for Direct CP Violation in K\u2192\u03c0\u03c0 Decay,''
  Phys.\ Rev.\ Lett.\  {\bf 115}, no. 21, 212001 (2015)
%  doi:10.1103/PhysRevLett.115.212001
  [\href{http://arXiv.org/abs/1505.07863}{arXiv:1505.07863}];
%\bibitem{Bai:2016ocm} 
%  Z.~Bai {\it et al.},
  %``Erratum: Standard-Model Prediction for Direct CP Violation in $K\to\pi\pi$ Decay,''
  Erratum, \href{http://arXiv.org/abs/1603.03065}{arXiv:1603.03065}.
  %%CITATION = ARXIV:1603.03065;%%

\bibitem{Czakon:2014xsa}
  M.~Czakon, P.~Fiedler and A.~Mitov,
  %``Resolving the Tevatron Top Quark Forward-Backward Asymmetry Puzzle: Fully Differential Next-to-Next-to-Leading-Order Calculation,''
  Phys.\ Rev.\ Lett.\  {\bf 115}, no. 5, 052001 (2015)
%  doi:10.1103/PhysRevLett.115.052001
  [\href{http://arXiv.org/abs/1411.3007}{arXiv:1411.3007}].
  %%CITATION = doi:10.1103/PhysRevLett.115.052001;%%

\bibitem{Isidori:2011qw}
  G.~Isidori, J.~F.~Kamenik, Z.~Ligeti and G.~Perez,
  %``Implications of the LHCb Evidence for Charm CP Violation,''
  Phys.\ Lett.\ B {\bf 711}, 46 (2012)
  [\href{http://arXiv.org/abs/1111.4987}{arXiv:1111.4987}].
  %%CITATION = ARXIV:1111.4987;%%

\bibitem{Gedalia:2012pi}
  O.~Gedalia, J.~F.~Kamenik, Z.~Ligeti and G.~Perez,
  %``On the Universality of CP Violation in Delta F = 1 Processes,''
  Phys.\ Lett.\ B {\bf 714}, 55 (2012)
  [\href{http://arXiv.org/abs/1202.5038}{arXiv:1202.5038}].
  %%CITATION = ARXIV:1202.5038;%%

\bibitem{Mahbubani:2012qq}
  R.~Mahbubani, M.~Papucci, G.~Perez, J.~T.~Ruderman and A.~Weiler,
  %``Light Nondegenerate Squarks at the LHC,''
  Phys.\ Rev.\ Lett.\  {\bf 110}, no. 15, 151804 (2013)
  [\href{http://arXiv.org/abs/1212.3328}{arXiv:1212.3328}].
  %%CITATION = ARXIV:1212.3328;%%

\bibitem{Nir:1993mx}
  Y.~Nir and N.~Seiberg,
  %``Should squarks be degenerate?,''
  Phys.\ Lett.\ B {\bf 309}, 337 (1993)
%  doi:10.1016/0370-2693(93)90942-B
  [\href{http://arXiv.org/abs/hep-ph/9304307}{hep-ph/9304307}].
  %%CITATION = doi:10.1016/0370-2693(93)90942-B;%%

\bibitem{Aaij:2015tna}
  R.~Aaij {\it et al.} [LHCb Collaboration],
  %``Search for hidden-sector bosons in $B^0 \!\to K^{*0}\mu^+\mu^-$ decays,''
  Phys.\ Rev.\ Lett.\  {\bf 115}, no. 16, 161802 (2015)
  %doi:10.1103/PhysRevLett.115.161802
  [\href{http://arXiv.org/abs/1508.04094}{arXiv:1508.04094}].
  %%CITATION = doi:10.1103/PhysRevLett.115.161802;%%

\bibitem{Freytsis:2010ne} 
  M.~Freytsis and Z.~Ligeti,
  %``On dark matter models with uniquely spin-dependent detection possibilities,''
  Phys.\ Rev.\ D {\bf 83}, 115009 (2011)
%  doi:10.1103/PhysRevD.83.115009
  [\href{http://arXiv.org/abs/1012.5317}{arXiv:1012.5317}].
  %%CITATION = doi:10.1103/PhysRevD.83.115009;%%

\bibitem{Freytsis:2009ct}
  M.~Freytsis, Z.~Ligeti and J.~Thaler,
  %``Constraining the Axion Portal with B ---> K l+ l-,''
  Phys.\ Rev.\ D {\bf 81}, 034001 (2010)
  [\href{http://arXiv.org/abs/0911.5355}{arXiv:0911.5355}].
  %%CITATION = ARXIV:0911.5355;%%

\bibitem{Godang:1997we}
  R.~Godang {\it et al.} [CLEO Collaboration],
  %``Observation of exclusive two-body B decays to kaons and pions,''
  Phys.\ Rev.\ Lett.\  {\bf 80}, 3456 (1998)
%  doi:10.1103/PhysRevLett.80.3456
  [\href{http://arXiv.org/abs/hep-ex/9711010}{hep-ex/9711010}].
  %%CITATION = doi:10.1103/PhysRevLett.80.3456;%%

\bibitem{Grossman:2002bu}
  Y.~Grossman, A.~L.~Kagan and Z.~Ligeti,
  %``Can the CP asymmetries in B ---> psi K(s) and B ---> psi K(L) differ?,''
  Phys.\ Lett.\ B {\bf 538}, 327 (2002)
%  doi:10.1016/S0370-2693(02)02027-0
  [\href{http://arXiv.org/abs/hep-ph/0204212}{hep-ph/0204212}].
  %%CITATION = doi:10.1016/S0370-2693(02)02027-0;%%

\bibitem{Jung:2012mp}
  M.~Jung,
  %``Determining weak phases from $B\to J/\psi P$ decays,''
  Phys.\ Rev.\ D {\bf 86}, 053008 (2012)
  [\href{http://arXiv.org/abs/1206.2050}{arXiv:1206.2050}].
  %%CITATION = ARXIV:1206.2050;%%

\bibitem{Frings:2015eva}
  P.~Frings, U.~Nierste and M.~Wiebusch,
  %``Penguin contributions to CP phases in $B_{d,s}$ decays to charmonium,''
  Phys.\ Rev.\ Lett.\  {\bf 115}, 061802 (2015)
  [\href{http://arXiv.org/abs/1503.00859}{arXiv:1503.00859}].
  %%CITATION = ARXIV:1503.00859;%%

\bibitem{Ligeti:2015yma}
  Z.~Ligeti and D.~J.~Robinson,
  %``Towards more precise determinations of the quark mixing phase $\beta$,''
  Phys.\ Rev.\ Lett.\  {\bf 115}, no. 25, 251801 (2015)
%  doi:10.1103/PhysRevLett.115.251801
  [\href{http://arXiv.org/abs/1507.06671}{arXiv:1507.06671}].
  %%CITATION = doi:10.1103/PhysRevLett.115.251801;%%

\bibitem{Jung:2015yma}
  M.~Jung,
  %``Branching ratio measurements and isospin violation in B-meson decays,''
  Phys.\ Lett.\ B {\bf 753}, 187 (2016)
%  doi:10.1016/j.physletb.2015.12.024
  [\href{http://arXiv.org/abs/1510.03423}{arXiv:1510.03423}].

\bibitem{Charles:2013aka}
  J.~Charles, S.~Descotes-Genon, Z.~Ligeti, S.~Monteil, M.~Papucci and K.~Trabelsi,
  %``Future sensitivity to new physics in $B_d, B_s$, and K mixings,''
  Phys.\ Rev.\ D {\bf 89}, no. 3, 033016 (2014)
  [\href{http://arXiv.org/abs/1309.2293}{arXiv:1309.2293}].
  %%CITATION = ARXIV:1309.2293;%%

\bibitem{Falk:2001hx}
  A.~F.~Falk, Y.~Grossman, Z.~Ligeti and A.~A.~Petrov,
  %``SU(3) breaking and D0 - anti-D0 mixing,''
  Phys.\ Rev.\ D {\bf 65}, 054034 (2002)
%  doi:10.1103/PhysRevD.65.054034
  [\href{http://arXiv.org/abs/hep-ph/0110317}{hep-ph/0110317}];
  %%CITATION = doi:10.1103/PhysRevD.65.054034;%%
%\bibitem{Falk:2004wg}
  A.~F.~Falk, Y.~Grossman, Z.~Ligeti, Y.~Nir and A.~A.~Petrov,
  %``The D0 - anti-D0 mass difference from a dispersion relation,''
  Phys.\ Rev.\ D {\bf 69}, 114021 (2004)
%  doi:10.1103/PhysRevD.69.114021
  [\href{http://arXiv.org/abs/hep-ph/0402204}{hep-ph/0402204}].
  %%CITATION = doi:10.1103/PhysRevD.69.114021;%%

\bibitem{Ishiwata:2015cga}
  K.~Ishiwata, Z.~Ligeti and M.~B.~Wise,
  %``New Vector-Like Fermions and Flavor Physics,''
  JHEP {\bf 1510}, 027 (2015)
%  doi:10.1007/JHEP10(2015)027
  [\href{http://arXiv.org/abs/1506.03484}{arXiv:1506.03484}].
  %%CITATION = doi:10.1007/JHEP10(2015)027;%%

\bibitem{AMtalk}
A.~Meyer, talk at this conference,
\href{http://indico.cern.ch/event/325831/contributions/757774/}
{http://indico.cern.ch/event/325831/contributions/757774/}.

\bibitem{Fox:2007in}
  P.~J.~Fox, Z.~Ligeti, M.~Papucci, G.~Perez and M.~D.~Schwartz,
  %``Deciphering top flavor violation at the LHC with $B$ factories,''
  Phys.\ Rev.\ D {\bf 78}, 054008 (2008)
%  doi:10.1103/PhysRevD.78.054008
  [\href{http://arXiv.org/abs/0704.1482}{arXiv:0704.1482}].
  %%CITATION = doi:10.1103/PhysRevD.78.054008;%%

\bibitem{Nir:2016zkd}
  Y.~Nir,
  %``Flavour Physics and CP Violation,''
  CERN-2015-001, pp.123-156
%  doi:10.5170/CERN-2015-001.123
  [\href{http://arXiv.org/abs/1605.00433}{arXiv:1605.00433}].

\bibitem{YStalk}
Y.~Sakai, talk at this conference,
\href{http://indico.cern.ch/event/325831/contributions/757784/}
{http://indico.cern.ch/event/325831/contributions/757784/}.

\bibitem{Bediaga:2012py} 
  R.~Aaij {\it et al.} [LHCb Collaboration],
  %``Implications of LHCb measurements and future prospects,''
  Eur.\ Phys.\ J.\ C {\bf 73}, no. 4, 2373 (2013)
%  doi:10.1140/epjc/s10052-013-2373-2
  [\href{http://arXiv.org/abs/1208.3355}{arXiv:1208.3355}].
  %%CITATION = doi:10.1140/epjc/s10052-013-2373-2;%%

\bibitem{Butler:2013kdw} 
  J.~N.~Butler {\it et al.} [Quark Flavor Physics Working Group Collaboration],
  %``Working Group Report: Quark Flavor Physics,''
  \href{http://arXiv.org/abs/1311.1076}{arXiv:1311.1076}.
  %%CITATION = ARXIV:1311.1076;%%

\bibitem{gammaTH}
A.~F.~Falk, private communications, unpublished (2001);\\
%\bibitem{Brod:2013sga} 
  J.~Brod and J.~Zupan,
  %``The ultimate theoretical error on $\gamma$ from $B \to DK$ decays,''
  JHEP {\bf 1401}, 051 (2014)
%  doi:10.1007/JHEP01(2014)051
  [\href{http://arXiv.org/abs/1308.5663}{arXiv:1308.5663}].
  %%CITATION = doi:10.1007/JHEP01(2014)051;%%

\bibitem{Laplace:2002ik}
  S.~Laplace, Z.~Ligeti, Y.~Nir and G.~Perez,
  %``Implications of the CP asymmetry in semileptonic B decay,''
  Phys.\ Rev.\ D {\bf 65}, 094040 (2002)
%  doi:10.1103/PhysRevD.65.094040
  [\href{http://arXiv.org/abs/hep-ph/0202010}{hep-ph/0202010}].
  %%CITATION = doi:10.1103/PhysRevD.65.094040;%%

\bibitem{Lenz:2011ti}
  A.~Lenz and U.~Nierste,
  %``Numerical Updates of Lifetimes and Mixing Parameters of B Mesons,''
  \href{http://arXiv.org/abs/1102.4274}{arXiv:1102.4274}.
  %%CITATION = ARXIV:1102.4274;%%

\bibitem{BGckm06}
B.~Grinstein, Plenary talk at CKM 2006,
\href{http://ckm2006.hepl.phys.nagoya-u.ac.jp/slide/Plenary.html}
{http://ckm2006.hepl.phys.nagoya-u.ac.jp/slide/Plenary.html}.


\end{thebibliography}
\end{document}